\documentstyle[preprint,prc,aps,floats,epsf]{revtex}

\tightenlines

\newcommand{\beq}{\begin{equation}}
\newcommand{\eeq}{\end{equation}}
\newcommand{\be}{\begin{eqnarray}}
\newcommand{\ee}{\end{eqnarray}}
\newcommand{\ben}{\begin{eqnarray*}}
\newcommand{\een}{\end{eqnarray*}}

\def\simgt{\stackrel{>}{{}_\sim}}

\newcommand{\lnablasq}{\overleftarrow{\nabla}{}^{\!2}}
\newcommand{\rnablasq}{\overrightarrow{\nabla}{}^{\!2}}
\newcommand{\wt}{\widetilde}
\newcommand{\Mbar}{\frac{M}{4\pi}(\mu+ip)}
\newcommand{\rz}{R}
\newcommand{\Real}{\mbox{Re}\,}
\newcommand{\Imag}{\mbox{Im}\,}
\newcommand{\sing}{$^1\!S_0$ }
\newcommand{\trip}{$^3\!S_1$ }
\newcommand{\LambdaNN}{\Lambda_{{\rm NN}}}
\newcommand{\order}{{\cal O}}

\begin{document}

\title{Removing Pions from Two-Nucleon Effective Field Theory}

\author{James~V.~Steele and R.~J.~Furnstahl}
\address{Department of Physics \\
         The Ohio State University,\ \ Columbus, OH\ \ 43210}

\date{August, 1998}

\maketitle

\begin{abstract}
Two-nucleon effective field theory (EFT)
beyond momenta of order the pion mass is studied for
both cutoff regularization and dimensional regularization
with power divergence subtraction (PDS).
Models with two mass scales illustrate how
effects of long-distance pion physics must be removed from the
coefficients that encode short-distance physics.
The modified effective range expansion shows that
treating pions perturbatively, as in the PDS power counting,
limits the radius of convergence of the EFT.
Predictions from both regularization
schemes with one-pion contributions are
compared to data.
The breakdown of the effective field theory
occurs for momenta of order $300\,$MeV, even using the modified
effective range expansion.
This behavior
is shown to be consistent with that expected from two-pion contributions.
\end{abstract}


\thispagestyle{empty}

\newpage


\section{Introduction}

Nucleon-nucleon effective field theory (NN EFT)
without pions, which applies at low energies, is now well understood.
Predictions are made by constructing the most
general Lagrangian with nucleon fields only
and fixing the coefficients by matching
calculated and experimental observables in a momentum expansion.
A key feature of NN EFT is the need for a nonperturbative treatment,
as implied by the existence of shallow bound states such as the deuteron.
Heavy
propagating nucleons lead to an enhancement of loop graphs \cite{weinberg},
which destroys the systematic power counting
characteristic of perturbative EFT's \cite{Ecker}.

Weinberg's solution was to sum certain diagrams to all
orders in the loop expansion
via the Schr\"odinger equation \cite{weinberg,biraold}.
Recent analyses find this summation to be problematic for systems with large 
scattering lengths \cite{KSW1,Maryland,Richardson}, which 
has led to alternative prescriptions 
within various regularization schemes \cite{PDS,Lepage,gege}.
The use of a given regularization scheme has implications
on the renormalization scheme chosen because in practice 
calculations are truncated.
The subsequent calculations of low-energy two-nucleon amplitudes have been 
shown to be 
systematic and regularization scheme independent \cite{bira,OurPaper,gege},
reproducing the effective range expansion of nonrelativistic
potential scattering \cite{bira}.
The unambiguous predictions of additional observables
are in good agreement with data, as seen, for
example, in the triplet state of threshold $nd$ scattering
\cite{bedaq}, which is dominated by two-body scattering.

The characteristic breakdown scale, or radius of convergence,
for a momentum expansion of the low-energy NN 
$S$-matrix is dictated by the pion.
In conventional effective range theory,
this breakdown is manifested by a branch point in $p\cot\delta$
at $p=m_\pi/2$ from the one-pion-exchange Yukawa potential (see Sec.~II).
More generally, the physical picture from effective field theory
anticipates a breakdown for momenta
of order $p\sim m_\pi$, as details comparable to 
the Compton wavelength of the pion are resolved. 
Predictions for external momenta above this point require
taking the pion explicitly into account.
In this paper, we address some of the open questions
about power counting with pions and the radius of
convergence of the NN EFT.

In Ref.~\cite{PDS}, a power counting scheme  was
introduced, in which all pion contributions to the scattering
amplitude enter perturbatively.
Calculations were simplified by the use of dimensional
regularization with power divergence subtraction (PDS).
In contrast to Weinberg's scheme,
this new approach sums 
only the leading contact interaction between nucleons to all
orders.

Forcing a summation of higher-order contact interactions can be
devastating in some regularization schemes --- such as dimensional
regularization with minimal subtraction --- leading to
undesirable systematics for a large scattering length
\cite{KSW1,OurPaper,Park}. 
PDS was introduced to define a consistent power counting scheme that
avoids this forced summation.
However, with a regularization that is solved numerically,
such as cutoff regularization,
it is simpler in practice to sum up these higher-order terms.
We have shown previously that regularization
schemes exhibiting naturalness in the bare parameters
have the proper EFT behavior even when higher-order terms are summed
\cite{OurPaper}.
In other words,
as long as the cutoff is suitably chosen so that the theory is 
natural (bare parameters of order unity),
the summation of higher-order terms does not affect the power
counting \cite{OurPaper}, implying that cutoff regularization 
is an equally valid scheme for calculations.

Properly adding the pion should, in principle, increase
the range in momentum for which the EFT is predictive
to the mass of particles not explicitly taken into account in the
Lagrangian.
Optimistic estimates in the literature anticipate a breakdown
on the order of the $\rho$ mass or $p\sim 800\,$MeV \cite{KSW1,Lepage,gege}.
However, both dimensional \cite{PDS} and cutoff \cite{Lepage} regularization
calculations of NN scattering with pions imply a breakdown
closer to $p\sim 300\,$MeV.
The subsequent application of EFT techniques to nuclear
matter could be problematic due to the close proximity of the Fermi
momentum ($280\,$MeV) to the breakdown point.

The observed breakdown of NN EFT could have various sources.
Explicit inclusion of correlated two-pion exchange, 
pion production, or 
the $\Delta(1232)$
may be necessary to increase the radius of convergence 
beyond several hundred MeV.
The breakdown scale could even be $m_\rho/2$. 
However, we 
should first establish that the EFT is indeed limited by physics
and not by how it is implemented.
In the cutoff approach, 
simply adding one-pion exchange to the
effective potential might not completely remove its influence on
the radius of convergence.
Qualitatively different reasons may limit the PDS approach.
A renormalization group analysis \cite{PDS} showed that PDS power counting
should break down at a new scale $\LambdaNN\sim 300\,$MeV,  
but there has been no systematic error analysis for PDS with
perturbative pions.

In this paper, we address these implementation issues.
Our approach generalizes the procedure of Ref.~\cite{OurPaper}
by requiring the
consistent removal of long-distance (pion) physics from
the coefficients that encode short-distance physics.
In Sec.~II,
we illustrate a technique to do this, called the modified effective
range expansion \cite{ModER}, which treats pions nonperturbatively.
We demonstrate its effectiveness and relevance
within the context of two toy models using cutoff
regularization. 
The modified effective range expansion is used in Sec.~III to show that
treating pions perturbatively within the PDS scheme 
limits the radius of convergence of the EFT.
In Sec.~IV, we compare NN scattering data to
results from applying these techniques to both regularization schemes.
The breakdown around $p\sim 300\,$MeV persists, but this scale
is shown to be consistent with that expected from two-pion contributions.
Section~V is a summary of our results.

\section{Examining the Need for the\\Modified Effective Range Expansion}

The effective range (ER) expansion for nonrelativistic NN scattering
\cite{Goldberger} is derived by using analyticity
arguments to relate the 
momentum times the cotangent of the phase shift
$p\cot\delta$ to an expansion in energy $E=p^2/M$, with $M$ the
nucleon mass,\footnote{We restrict the
discussion throughout this paper to interactions 
in the center of mass between two nucleons
with a reduced mass $M/2$, $M=940\,$MeV, and neglect electromagnetic
effects.  We also take 
$m_\pi=140\,$MeV, $m_\rho=770\,$MeV, $f_\pi = 92\,$MeV, and
$\alpha_\pi=0.075$ below.}
\beq
p\cot\delta = -\frac1{a_s} + \frac12\, p^2
\sum_{n=0}^\infty r_n \frac{p^{2n}}{\Lambda^{2n}} \, .
\label{ER}
\eeq
Here the expansion is written in terms of 
the lightest mass scale $\Lambda$ that characterizes the
underlying dynamics \cite{PDS}.
For example,
a Yukawa or exponential potential with mass $\Lambda$ limits the
region of analyticity and hence validity of Eq.~(\ref{ER}) to
$p\le \Lambda/2$ \cite{Newton}.
Although the scattering
length can take on any value,
we expect the ranges $r_n$ to be of order $1/\Lambda$
\cite{Goldberger}; this is  empirically shown below.
Special considerations
are needed for the behavior of these ranges when the underlying
potential is strong, as discussed in Sec.~IV.

In our earlier work \cite{OurPaper}, we used a short-range delta-shell
potential at $r=1/m_\rho$
in the radial Schr\"odinger equation as a clean laboratory
with which to compare various regularization schemes.
As each new order in the momentum expansion of the effective
potential was matched to data,
we verified the improvement of the error
\beq
\Delta(p\cot\delta)\equiv
|p\cot\delta_{\rm eff}- p\cot\delta_{\rm true}|
\eeq
by powers of $p^2$.
Contributions from different orders in the $p^2$
expansion become comparable at what is 
known as the
radius of convergence of the effective field theory.
We expect this to occur for momenta comparable with
the underlying scale $\Lambda$. 
We demonstrated this by looking at error plots \cite{Lepage}
 of $\Delta(p\cot\delta)$
as a function of momentum.
Both cutoff and PDS regularizations were found to give systematic behavior
independent of the details of the underlying theory.

\subsection{Delta-Shell and Exponential Model}

We now add a long-range part to the underlying toy theory to model the
inclusion of the pion in the analysis of NN scattering.
We first study the addition of an
exponential well \cite{Newton} with mass $m_\pi$,
\beq
V(r)=-g \frac{m_\rho}{M}\; \delta(r-1/m_\rho) - V_0 \; e^{-m_\pi r} \ .
\label{toy}
\eeq
The exponential and delta-shell
model the long- and short-range potentials respectively.
Our ability to solve Eq.~(\ref{toy}) analytically
allows us to validate the numerical procedures used below in more
realistic cases.
The exponential results in singularities in
$p\cot\delta$ starting at $p=-im_\pi/2$ analogous to the branch points
from a Yukawa \cite{Newton}.
We use a ``chiral-symmetry'' coupling $V_0=m_\pi^2/M$.
The delta-shell coupling
will be taken to be $g=1.01$ for comparison to our earlier
paper \cite{OurPaper}.

We determine the phase shift for $V(r)$ by calculating the $S$-wave
Jost function ${\cal F}(p)$, which plays a key role in the
modified effective range expansion \cite{ModER}.
This is accomplished by solving the Schr\"odinger equation
for the Jost solution $f(p,r)$,
\beq
\left[ \frac{d^2}{dr^2} + p^2 - MV(r) \right] f(p,r) = 0\ .
\eeq
The Jost solution satisfies the asymptotic
condition $f(p,r\to\infty)\to e^{ipr}$
and has the property $f(p,0)={\cal F}(p)$.
The exact Jost function for an exponential
potential \cite{Newton} can be generalized for the potential in Eq.~(\ref{toy})
and written in terms of Bessel and gamma functions,
\beq
{\cal F}(p) =  \left(\frac{y_0}{2} \right)^{\!\!\nu} \Gamma(1-\nu)
\left[ J_{-\nu}(y_0) + \frac{g\pi}{\sin\pi\nu} \frac{m_\rho}{m_\pi}
J_{-\nu}(y) \left( J_\nu(y)
J_{-\nu}(y_0) - J_{-\nu}(y) J_{\nu}(y_0) \right) \right] \ ,
\label{jost}
\eeq
with $\nu=2ip/m_\pi$, $y_0=2\sqrt{MV_0}/m_\pi$, and
$y=y_0 e^{-m_\pi/2m_\rho}$.
This satisfies the well known properties of the Jost
function, namely ${\cal F}(p)={\cal F}^*(-p^*)$
and for $V\to0$ or $p\to\infty$,
${\cal F}(p)\to1$.
We can extract the exact phase shift $\delta_{\rm true}$ by using
${\cal F}(p)=|{\cal F}(p)| e^{-i\delta_{\rm true}}$.

Expressing the Bessel functions in Eq.~(\ref{jost}) as a series, 
the Jost function can be expanded in powers of $1/m_\rho$,
\be
{\cal F}(p) = \left(1-g- g \frac{ip}{m_\rho} \right)
{\cal J}_{-\nu}(y_0)  + g \frac{y_0 m_\pi}{2 m_\rho}{\cal J}_{-\nu}'(y_0) +
\order\!\left(\frac{Q^2}{m_\rho^2}\right) \, ,
\label{series}
\ee
with $Q=\{m_\pi,p\}$ and
\be
{\cal J}_{-\nu}(y_0) \equiv 
\sum_{k=0}^\infty \frac{(-1)^k \; (y_0/2)^{2k}}{k! \,
(1-\nu) \ldots (k-\nu)} \, .
\label{series2}
\ee
Equation~(\ref{series2}) shows
explicitly the singularities starting at $p=-im_\pi/2$
that limit the region of analyticity of the effective range
expansion.
These singularities always enter multiplied by the strength of the
potential $y_0$ and have less impact on the radius of convergence
of the EFT as $y_0$ decreases (see below for examples).

Based on the above discussion,
using a purely short-range effective potential to describe $V(r)$
should result in a breakdown at $p\sim m_\pi/2$.
For $S$-waves, the effective potential is
\be
V_{\rm eff}(r) = C_0 \delta^3(r) - \frac12 C_2 \left( 
\lnablasq \delta^3(r) + \delta^3(r) \rnablasq \right) + \ldots \ ,
\label{Veffr}
\ee
which in momentum space becomes
\be
V_{\rm eff}(p,p') = C_0 + C_2 \frac{p^2+p'{}^2}2 + \ldots \ . 
\label{Veff}
\ee
It can be fixed to a given order in $p$ by matching the $C_{2n}$'s to the ER
parameters, Eq.~(\ref{ER}).
We do this by guessing an initial value for each of the $C_{2n}$'s
and calculating $p\cot\delta_{\rm eff}$.
Then a fit of the difference
\beq
\Delta (p\cot\delta) = \alpha + \beta\, \frac{p^2}{\Lambda^2}
+ \ldots \ ,
\label{fit}
\eeq
to a polynomial in $p^2/\Lambda^2$ results in the constants
$\alpha,\beta,\ldots$,
which are minimized up to the order of $V_{\rm eff}(r)$
with respect to variations in the
$C_{2n}$'s~\cite{OurPaper}.
Weighting the fit by both the expected theoretical error and
experimental uncertainty ensures that the EFT constants are properly
obtained.
The singular nature of $V_{\rm eff}(r)$ requires regularization.
We concentrate on using cutoff regularization with a cutoff
$\Lambda_c$ in this section,
as detailed in Ref.~\cite{OurPaper}.
We focus on PDS in the next section.

\begin{figure}[t]
\begin{center}
\leavevmode
\hbox{
\hspace{-.5cm}
\epsfxsize=3.4in
\epsffile{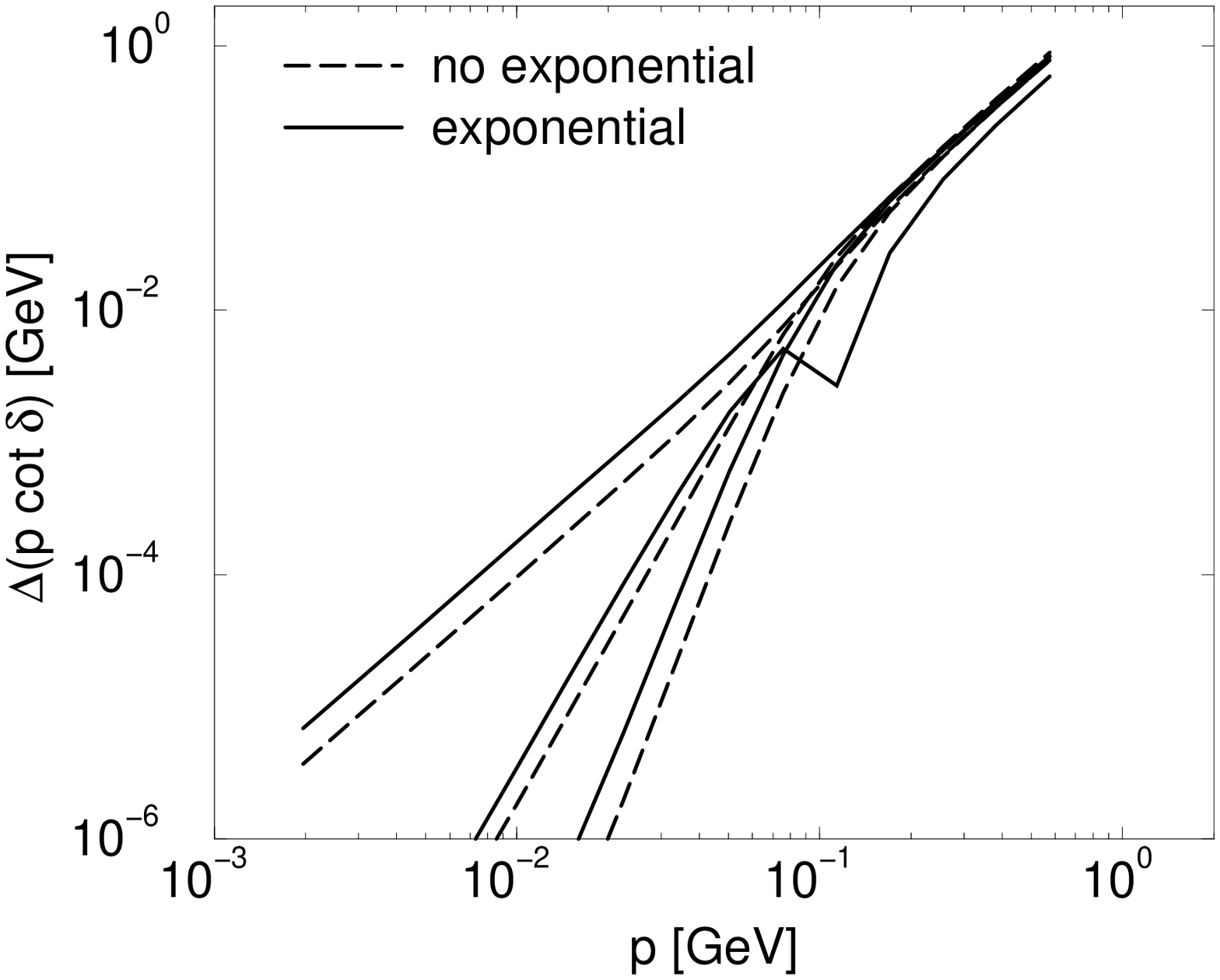}
\hspace{-.4cm}
\epsfxsize=3.4in
\epsffile{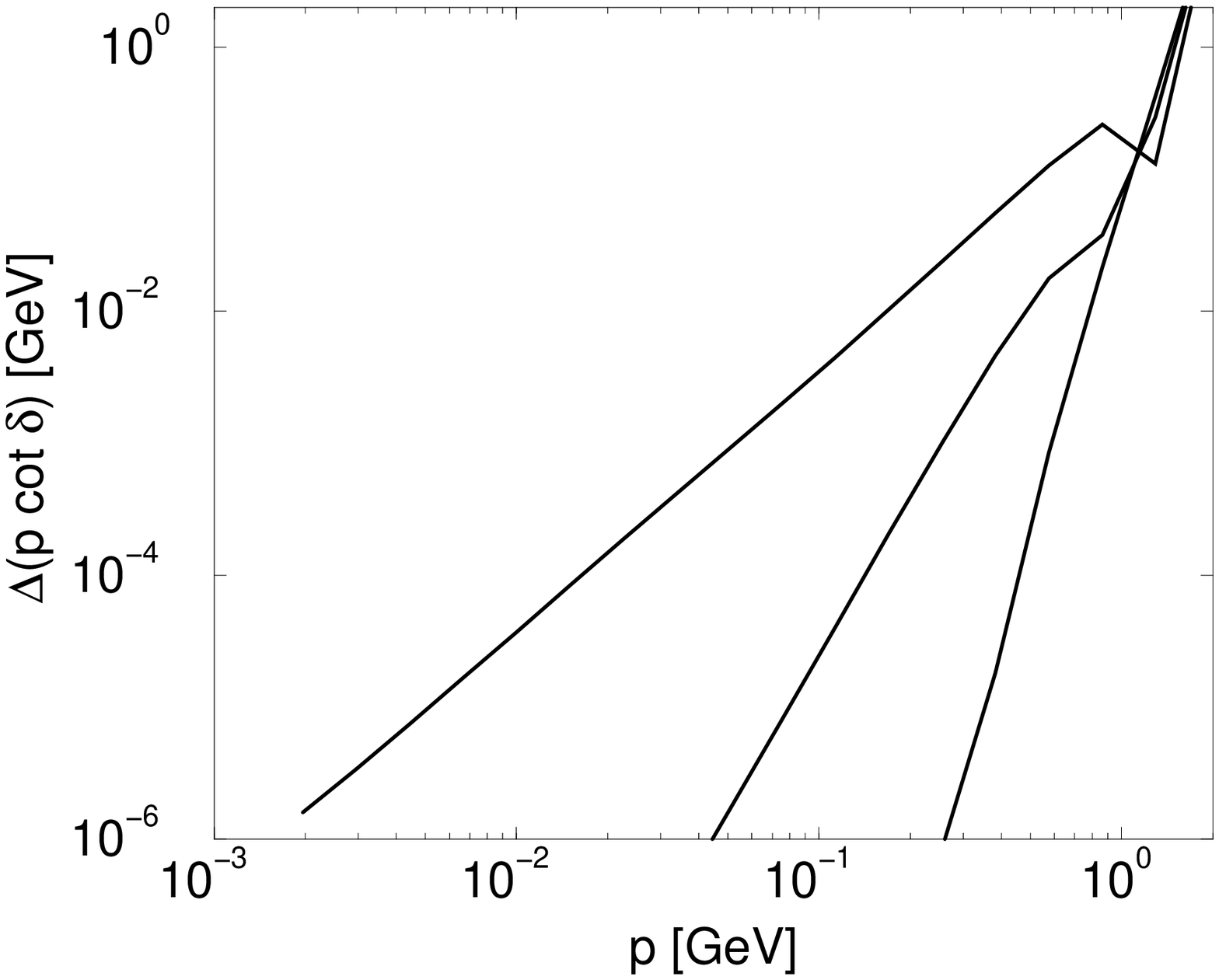}
}
\end{center}
\caption{\label{exponent} The model of a delta-shell plus exponential
potential, Eq.~(\protect\ref{toy}), with low energy constants fit by
using the conventional effective range expansion (left) and modified
effective range expansion (right).
The radius of convergence on the left is about $90\,$MeV regardless of
whether the exponential potential is explicitly included (solid) or not 
included (dashed) in the effective potential.
Using the modified effective range expansion 
increases the radius of convergence to about $1\,$GeV.}
\vspace{.3cm}
\end{figure}

The cutoff $\Lambda_c$ is roughly adjusted to minimize the error,
which identifies the resolution scale of the underlying short-distance
physics \cite{Lepage}. 
The results for using a short-distance effective potential to describe
Eq.~(\ref{toy})
are only weakly dependent on $\Lambda_c$, but are best for 
$\Lambda_c\sim70\,$MeV, which are shown as the
dashed lines in the left plot of Fig.~\ref{exponent}.
We plot the error $\Delta(p\cot\delta)$ as a function of momentum on a
log-log plot.
The lines with increasing slope refer to one, two, and three constants
fixed in $V_{\rm eff}$ respectively.
The radius of convergence can be read from the graph to be about
$90\,$MeV.
This breakdown point is directly connected to the singularity in $p\cot\delta$
at $m_\pi/2$, as can be confirmed by varying the mass in the
exponential of Eq.~(\ref{toy}).

It might be expected
that by including the long-range behavior of the exponential potential $-V_0
e^{-m_\pi r}$ explicitly in
the effective potential Eq.~(\ref{Veffr}), we can increase the radius
of convergence
beyond $m_\pi/2$.
Although the values of the $C_{2n}$'s from matching to the ER parameters
in this case change slightly, the subsequent error plots
are almost unchanged, as shown by the
solid lines in the left plot of Fig.~\ref{exponent}.
{\it Naively taking the long-range physics into account in the cutoff 
effective
field theory is simply not enough to yield an improved result.\/}
A nonperturbative treatment of long-distance physics with a cutoff
leads to contamination in the short-distance coefficients.
This can be fixed by appropriate counterterms as discussed in
Sec.~III.C. 
But first, we show how to fix this problem using the modified ER
expansion.

\subsection{The Modified Effective Range Expansion}

Properly accounting for the long-distance physics (the
pion) is equivalent to removing its effects from the ER expansion.
This should result in a new function with a region of analyticity
larger than $p\cot\delta$ 
and lead to an improved radius of convergence for the EFT.
After separating a general potential into a long- and short-range part
$V=V_L+V_S$, such a function can be constructed and
expressed in terms of the full phase shift $\delta$ and the
Jost function ${\cal F}_L(p)$, Jost solution
$f_L(p,r)$, and phase shift $\delta_L$ 
for the long-range potential {\it only\/} \cite{ModER}:
\beq
K(p) = \frac{p\cot(\delta-\delta_L)}{|{\cal F}_L(p)|^2}
+ \Real \left[ \frac{f_L'(p,0)}{{\cal F}_L(p)} \right] \equiv
-\frac1{\wt{a}_s} + \frac12 p^2 \sum_{n=0}^\infty \wt{r}_n
\frac{p^{2n}}{\wt{\Lambda}^{2n}} \ .
\label{MERF}
\eeq
The prime denotes differentiation with respect to $r$.
Equation~(\ref{MERF}), which is called the modified effective range
expansion, 
has been most commonly used to remove Coulomb effects
from the analysis of NN scattering; it 
can also be generalized to higher angular momenta $l$ \cite{ModER}.
The combination $K(p)$ removes the effect of the long-range potential $V_L$, 
increasing the relevant $\wt{\Lambda}$
to the order of the short-range potential.
Thus we expect that $\wt{r}_n \sim 1/\wt{\Lambda}$.
For the toy theory of Eq.~(\ref{toy}), this would mean
removing the pion singularity and moving the radius of convergence
up to $m_\rho$, the scale of
the delta-shell.
Note that when the long-range potential is zero, $\delta_L=0$,
$f_L(p,r)=e^{ipr}$, and the conventional effective range expansion is
recovered.

We should emphasize that the reason we introduce the modified
ER expansion  is to
ensure that the addition of the long-range potential to our EFT does not
contaminate
the process of fitting the low energy constants.
Once the constants are fit properly, we can return to the evaluation
of the familiar observable $p\cot\delta$ to assess the improvements
made by the effective field theory.
When the phase shift is close to zero, $p\cot\delta$ becomes abnormally large 
and it can be advantageous to look at error plots of $K(p)$ instead.

We now apply the modified ER procedure to the toy model of Eq.~(\ref{toy}).
We again include the exponential $-V_0 e^{-m_\pi r}$ 
in the EFT, but now match the low-energy
constants $C_{2n}$ by enforcing $\Delta K(p)=0$ to the order in $p^2$ we are
working, similar to what was done for $\Delta(p \cot\delta)$
in Eq.~(\ref{fit}).
The exact Jost function Eq.~(\ref{jost}) with $g=0$ is the
long-distance Jost function ${\cal F}_L(p)$ 
used in Eq.~(\ref{MERF}).
Since both the EFT and the true theory have the same long-distance
behavior (by construction), 
the term in the modified ER function that depends on $f_L'(p,0)$
cancels in the difference.

The results of the modified ER fit are shown in the right plot of
Fig.~\ref{exponent}.
The optimal (natural) value of $\Lambda_c$ is now about $800\,$MeV 
and the radius of convergence is seen to be about $1\,$GeV.
Furthermore, the EFT constants that are matched using the modified ER 
procedure agree exactly
with the values from the delta-shell potential alone
\cite{OurPaper}.
This verifies that the long-distance physics is successfully removed.
For more general potentials, 
with long- and short-distance scales of $m_\pi$ and
$m_\rho$ respectively, it can be shown that the modified ER expansion
Eq.~(\ref{MERF}) is the
same as the pure short-distance ER expansion up to terms of 
$\order(m_\pi^2/m_\rho^2)$ \cite{PhysRep}.
These corrections are actually zero for a delta-shell potential.

\begin{figure}
\vspace{-.2in}
\begin{center}
\leavevmode
\epsfxsize=3.4in
\epsffile{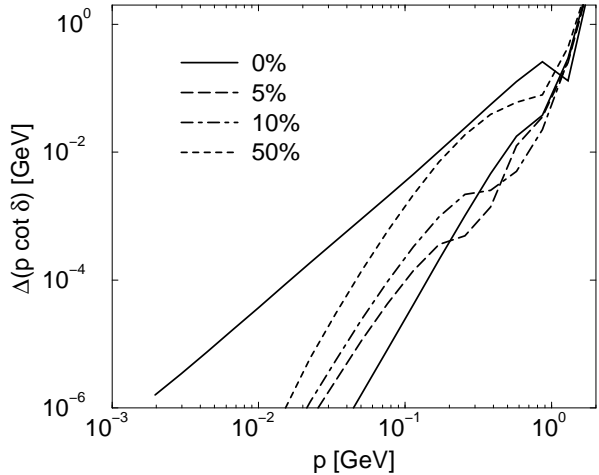}
\end{center}
\caption{\label{error} The dependence of the
two-constant fit to Eq.~(\protect\ref{toy})
on the percentage error in $V_0$.}
\end{figure}

When comparing to data, we expect
to have some error in the long-range potential used in our
effective theory compared to the underlying theory.
We can estimate the effect this will have on our results
by purposely detuning the
value of $V_0$ used in the long-range part of our effective potential.
The fit to three constants is sensitive enough to fail if the error is
larger than a fraction of a percent.
The fit to one constant gives consistent results regardless of the error.
The comparison of the results for a fit to two constants is shown in
Fig.~\ref{error} for errors of $5$, $10$, and $50$ percent.
Although the results are not as clean as in the error-free case, we
can read off a radius of convergence by taking a straight line from
the low energy behavior.
This shows that the radius of convergence inferred from the
plot slowly reduces as the error in $V_0$ increases.
As long as the discrepancy in the long-range part of the effective
field theory and that from data do not differ by more than 5\%, the
results for one and two constants are fairly stable.

Using the modified ER expansion is a guaranteed way to properly 
incorporate the long-distance physics in the EFT.
However, at least some improvement in the radius of convergence 
was found in the case of NN scattering without this technique 
\cite{biraold,Lepage}.
We use a second model in the next subsection to further investigate
when the modified ER expansion is needed.

\subsection{Two-Yukawa Model}

A more representative toy model of nuclear forces 
consists of two Yukawas of masses
$m_\pi$ and $m_\rho$.
We require
\beq
V(r) \stackrel{r\to\infty}{\longrightarrow} -\alpha_\rho \,
\frac{e^{-m_\rho r}}r - \alpha_\pi \, \frac{e^{-m_\pi r}}r \, ,
\label{twoyuk}
\eeq
but for small $r$ the results are practically independent of whether
a cut-off Yukawa \cite{Lepage} or a hard core repulsion are used to
regulate the potential at the origin.
For simplicity, we take Eq.~(\ref{twoyuk}) as the exact potential and
merely exclude the origin.
The couplings are initially taken to be
$\alpha_\pi=0.075$, as given by one-pion exchange in \sing NN
scattering,
and  $\alpha_\rho=1.05$, which gives the large scattering length
found in that channel,
$a_s=-25\,$fm.
Equation~(\ref{twoyuk}) models the short-distance physics with a single scale.
Nuclear phenomenology suggests that the short-range Yukawa 
actually should be replaced by the combination of 
attractive  and repulsive Yukawas resulting from the exchange of
scalar and vector mesons \cite{machleidt}. 
We will come back to this issue
in Sec.~IV.

Since the long-distance physics is dictated by the pion, 
the effective Lagrangian that describes this model problem is
identical to that for NN scattering \cite{weinberg,PDS}
\be
{\cal L}_{\rm eff} &=& N^\dagger \left[ i \partial_t + \frac{\nabla^2}{2M}
-\frac{g_A}{2f_\pi} \sigma \cdot \nabla (\pi \cdot \tau) \right] N
\nonumber\\
&&- C_0 (N^\dagger N)^2 + \frac12 C_2 \left[ (N^\dagger
\nabla N)^2 + ((\nabla N^\dagger) N)^2 \right] + \ldots \, ,
\label{Leff}
\ee
with the ellipses representing terms with more than one pion field 
as well as higher-derivative terms.  
Neglecting retardation effects since they are absent from
the underlying toy model by construction,
Eq.~(\ref{Leff}) can be expressed in terms of a potential (for the
\sing channel)
\be
V_{\rm eff}(r) = - \alpha_\pi \frac{e^{-m_\pi r}}r
+ \left( C_0 + \frac{4\pi \alpha_\pi}{m_\pi^2}
\right) \delta^3(r) 
- \frac12 C_2 \left( \lnablasq
\delta^3(r) + \delta^3(r) \rnablasq \right) + \ldots \ ,
\label{Veff1}
\ee
with $\alpha_\pi=g_A^2 m_\pi^2/16\pi f_\pi^2$, showing
that the long-distance physics of the pion from Eq.~(\ref{twoyuk})
is included exactly.
The full effective potential, including the pion Yukawa,
is regularized at short distances
by a cutoff $\Lambda_c$ \cite{Lepage}. 

The Jost functions for both the full phase shift and the long-distance part
required in Eq.~(\ref{MERF}) are calculated numerically, 
as outlined in Appendix A.
Note that when calculating the modified ER expansion $K(p)$,
we still impose a cutoff on
the potential Eq.~(\ref{twoyuk}), as would be done if
the underlying potential were not exactly known.

\begin{figure}[t]
\vspace{-.2in}
\begin{center}
\leavevmode
\hbox{
\hspace{-.5cm}
\epsfxsize=3.4in
\epsffile{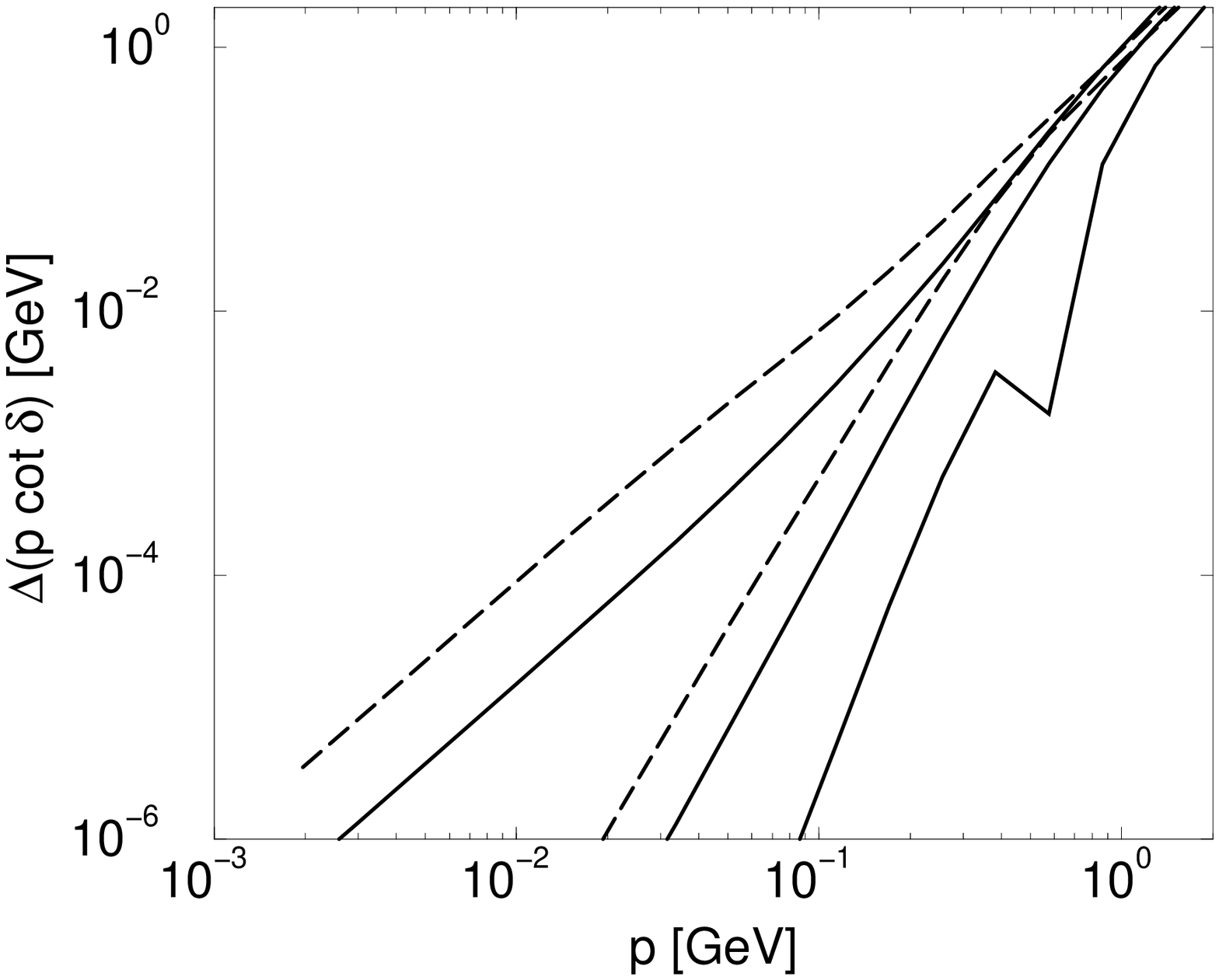}
\hspace{-.4cm}
\epsfxsize=3.4in
\epsffile{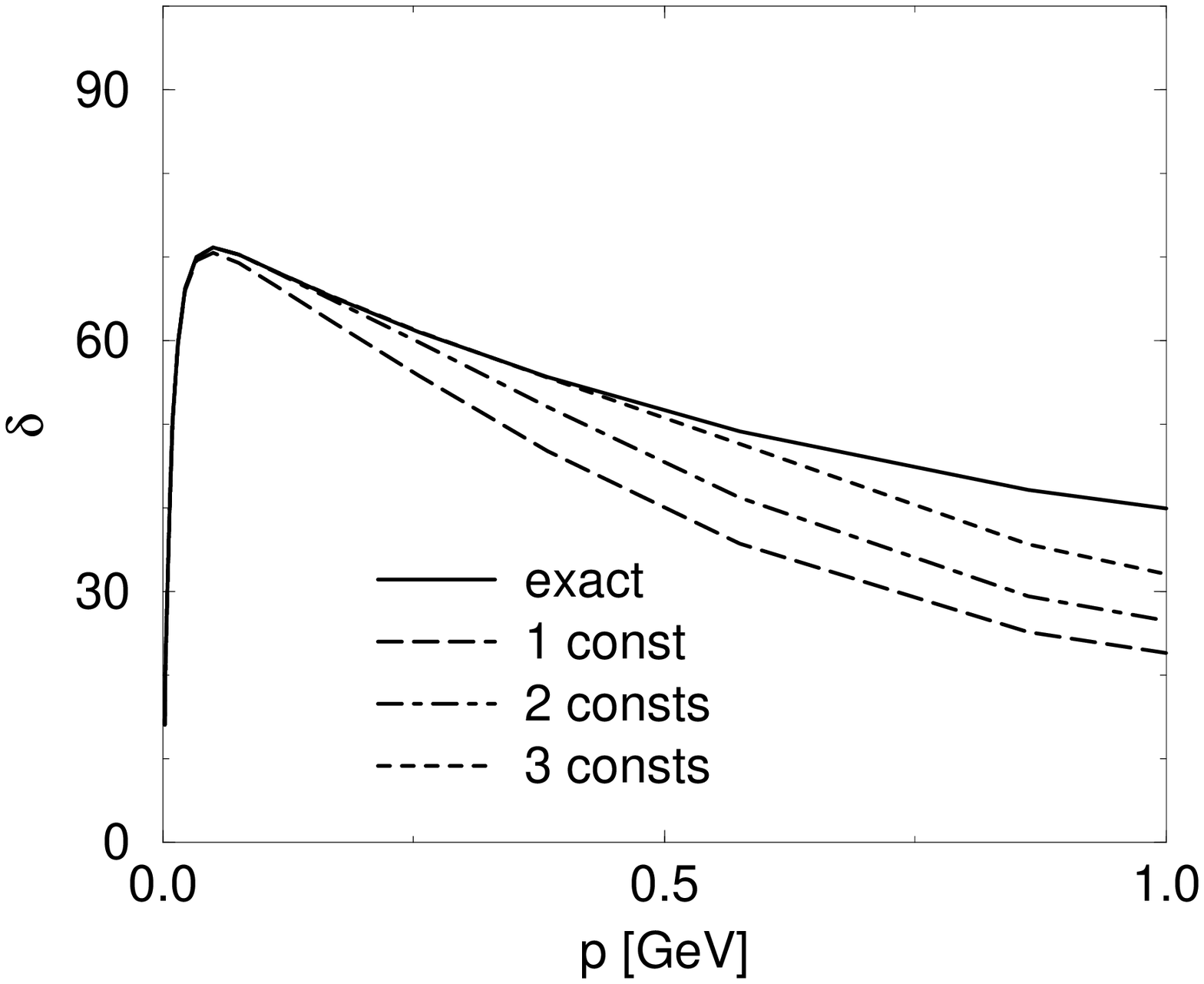}
}
\end{center}
\caption{\label{yukawa} The error in
$p\cot\delta$ (left) and the phase shift in degrees (right)
for the two-Yukawa toy model.
The solid curves on the left show errors from the modified ER
fit to one, two, or three constants,
which are independent of the strength of $V_L$.
Without using the modified ER procedure, similar results are obtained
for one and two constants with a weak $\alpha_\pi$. 
The dashed curves on the left
show the errors as well as the radius of convergence
worsen if $\alpha_\pi$ is made a factor of three stronger.}
\end{figure}

The results from using the modified ER expansion to fit one, two,
and three constants are shown by the solid
lines in the left plot of Fig.~\ref{yukawa} 
for the optimal value $\Lambda_c=600\,$MeV.
The phase shifts themselves are shown in the right plot.
While the improvement in the phase shifts
is manifest, it is clear that the error plots
are needed to make a quantitative assessment of the convergence. 
The plots show the radius of convergence is between $500$ and $600\,$MeV, 
which is comparable to
the range of analyticity $m_\rho/2$ of the modified ER expansion.
For comparison, we also used a delta-shell at $r=1/m_\rho$ as the
short-distance potential (see Appendix~A).
This gives a radius of convergence of about $1\,$GeV, which is in
agreement 
with the exponential potential model in the previous subsection,
and shows the different impact 
from branch point and pole singularities in $p\cot\delta$.

\begin{table}[t]
  \caption{\label{tab0}Conventional and modified effective range parameters
    for the two-Yukawa model. The pion Yukawa was cut off at
    $p=500\,$MeV to calculate the modified ER function.
    In parentheses the ranges are expressed in terms of 
    corresponding powers of 
    the expected underlying scale.}
\vspace{.5cm}
\begin{tabular}{c|ll@{\ }ll@{\ }ll@{\ }l}
 & \multicolumn{1}{c}{$a_s$ (fm)} & \multicolumn{2}{c}{$r_0$ (fm)} &
\multicolumn{2}{c}{$r_1/\Lambda^2$ (fm$^{3}$)}
& \multicolumn{2}{c}{$r_2/\Lambda^4$ (fm$^{5}$)} \\ \hline
ER & $-25.1$ & $1.63$ & ($1.2/m_\pi$) &
$-3.4$ &($-1.2/m_\pi^3$) & $16.$ & ($2.9/m_\pi^5$) \\
Mod. ER & $-1.72$ & $0.667$ & ($2.6/m_\rho$)
& $-0.024$ & ($-1.4/m_\rho^3$) & $0.0028$ &  ($2.6/m_\rho^5$) \\
\end{tabular}
\end{table}

The first few ER parameters from $p\cot\delta$ in Eq.~(\ref{ER})
are given in Table~\ref{tab0}.
As anticipated, the ranges are all on the order of $1/m_\pi$.
The modified ER expansion leads to values more naturally expressed in powers
of $1/m_\rho$.
Once again we find that the modified ER parameters truly have
the long-distance physics removed, allowing for a clean fit to our
EFT and resulting in an improved radius of convergence.
Introducing some error into the value of $\alpha_\pi$, as studied in the
last subsection, leads not only to an inability to fit three
constants but also to unnatural modified ER parameters.
This effect may provide a way to work towards
a more accurate value for $\alpha_\pi$.

Using the conventional ER expansion to fit one or two constants in the EFT, 
though, gives
results just as good as from using the modified ER expansion.
It only fails at three constants. 
Since
$\alpha_\pi\sim m_\pi/m_N$ is weak, the importance of the
long-distance potential, as well as the need for the modified ER
expansion, are reduced. 
The same result occurs in the model of the exponential well
if the strength of the potential is decreased. 
This behavior will be explained in the next section.
Regardless, the robustness of the fit to the ER parameters
for the two approaches is not qualitatively the same. 
The
success of the conventional ER fit depends strongly on the initial
guess for the EFT constants and the weighting of the fit in
Eq.~(\ref{fit}) by the theoretical error. 
These hindrances, which
do not play a major role in the modified ER procedure, could make it
difficult to fit to real data with experimental uncertainties.

By increasing $\alpha_\pi$, we should see the radius of convergence
from using the conventional ER fit decrease while the modified ER result
is unchanged.
This is illustrated by the dashed lines in the left plot of
Fig.~\ref{yukawa}, 
which show results for the conventional ER fit after
enhancing $\alpha_\pi$ by a factor of three.
The radius of convergence
is only about $300\,$MeV for the optimal $\Lambda_c=400\,$MeV.
The results from using the modified ER procedure
are insensitive to the strength of
the long-range coupling constant.

In summary, for a basic two-scale problem, the modified ER procedure
provides a
systematic removal of the long-distance physics, which ensures a proper,
robust fit to the low-energy constants, even with a strong long-range
potential.
This often results in a better radius of convergence for our EFT,
although the improvement is not significant for a weak long-range potential.
In addition, the modified ER parameters $\wt{r}_n$ from Eq.~(\ref{MERF})
are all on the order of the range of the short-distance potential,
$1/m_\rho$.

\section{Are Nonperturbative Pions Needed?}

We now turn to the PDS regularization scheme introduced in
Ref.~\cite{PDS}, which used the language of Feynman diagrams.
The nonrelativistic Lagrangian of Eq.~(\ref{Leff}) gives the
pion-nucleon vertex
\beq
\parbox{10mm}{
\epsffile{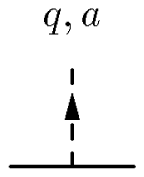}
}
\;=\; \frac{g_A}{2f_\pi} \tau^a\, \sigma \cdot q \, , 
\eeq
and so, after projection onto the \sing partial wave,
one-pion exchange contributes ($q=p-p'$)
\beq
\frac{g_A^2}{4f_\pi^2} \left[ -1 +
\frac{m_\pi^2}{4p^2} \,\ln\! \left( 1+ \frac{4p^2}{m_\pi^2} \right)
\right]
\label{OPE}
\eeq
to the full amplitude
\beq
{\cal A} = \frac{4\pi}{M} \frac{e^{2i\delta}-1}{2ip} 
         = \frac{4\pi}{M} \frac{1}{p\cot\delta - ip} \, .
\label{ampl}
\eeq
Notice that the amplitude develops a branch point in momentum at
$p=-i m_\pi/2$ when the pion Eq.~(\ref{OPE}) 
is included, as expected from the
effective range expansion.

The loop integrals are evaluated in dimensional regularization with
the PDS scheme, which subtracts all poles coming from four or fewer
dimensions \cite{PDS}.
For example,
the one-loop bubble diagram of two nucleons with external momentum
$p=\sqrt{ME}$ becomes
\beq
-i \left(\frac{\mu}2\right)^{\!\!4-D} \int\!\! \frac{d^Dq}{(2\pi)^D}\;
\frac{i}{E/2 + q_0 - {\bf q}^2/2M + i\epsilon}\;
\frac{i}{E/2 - q_0 - {\bf q}^2/2M + i\epsilon}
= - \Mbar\, .
\label{PDSint}
\eeq
An arbitrary mass scale $\mu$ is introduced to ensure that coupling
constants multiplying the integral Eq.~(\ref{PDSint})
retain their dimension
when $D$ is taken different from four.
Physical quantities are independent of $\mu$, which has two
consequences: (i) the coupling constants $C_{2n}$ in Eq.~(\ref{Leff})
must depend on $\mu$ to cancel the $\mu$-dependence of the loop
integrals;
and (ii) $\mu$ must be of the same order as $p$ or smaller
to ensure that the integral
Eq.~(\ref{PDSint}) is properly counted in a momentum expansion.

\subsection{PDS Renormalization Group}

\begin{figure}
\vspace{-.2in}
\begin{center}
\leavevmode
\epsfxsize=6in
\epsffile{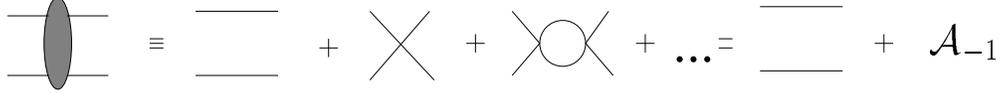}
\vspace{-.2in}
\end{center}
\caption{\label{feynm1} Leading order diagrams in the PDS
power counting with $C_0$ at each vertex.}
\end{figure}

The PDS power counting requires
the constants in the effective
Lagrangian  scale like $C_0\sim 1/\mu$ and 
$C_2  \sim 1/\mu^2$ 
when considering momenta $p \sim \mu \gg 1/a_s$ \cite{PDS}.
This is particularly relevant for
$S$-wave NN scattering, where the scattering length is large.
Furthermore, one-pion exchange scales like $\order(p^0)$ as in
Eq.~(\ref{OPE}) and 
nucleon loops scale like $\order(p)$
as in Eq.~(\ref{PDSint}).
Therefore, every $n$--loop diagram with insertions of the contact term $C_0$
scales like $\order(p^{-1})$.
To have a consistent power counting,
they must be summed \cite{PDS},
producing the leading-order amplitude 
depicted in Fig.~\ref{feynm1},
\beq
{\cal A}_{-1} = -\frac{C_0}{1+C_0 \Mbar}  \ .
\eeq
Matching ${\cal A}_{-1}$ to the leading-order
effective range expansion then determines the
$\mu$ dependence of $C_0$,
\beq
C_0(\mu) = \frac{4\pi}{M} \left( \frac1{-\mu + 1/a_s} \right) 
   + \order(\mu^0) \ ,
\label{C0m1}
\eeq
confirming the scaling $C_0\sim 1/\mu$.

\begin{figure}
\begin{center}
\leavevmode
\epsfxsize=4in
\epsffile{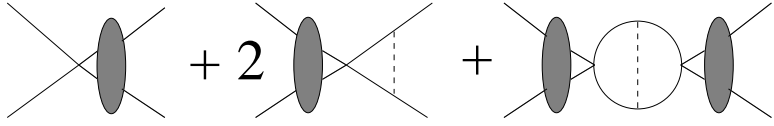}
\end{center}
\caption{\label{rgeqn} All $\mu$-dependent diagrams that contribute to
the $\beta$-function for $C_0$ to order $\order(p^0)$.}
\end{figure}

At subleading order $\order(p^0)$, perturbative corrections from
one-pion exchange and $C_2$ insertions begin to contribute.
For non-zero pion mass, a term of the form $D_2\, m_\pi^2$ also enters.
Since to this order only the combination $C_0+m_\pi^2 D_2$ 
appears, we absorb $D_2$ into the definition of $C_0$ for convenience.
Matching the amplitude to this order in the effective range expansion 
\beq
  p \cot\delta = ip + \frac{4\pi}{M{\cal A}_{-1}}
        - \frac{4\pi {\cal A}_0}{M{\cal A}_{-1}^2} + \cdots
\eeq
determines the $\mu$-dependence of the low-energy constants.
Using the explicit expression for the subleading amplitude ${\cal A}_0$ 
in Ref.~\cite{PDS}, we obtain
\be
C_0(\mu) &=& \frac{4\pi}{M} \left(
\frac{1+2 (\mu-m_\pi)/\LambdaNN}{-\mu_\pi+ 1/a_s} \right)
   + \order(\mu)  \, \label{C00} \\[2pt]
C_2(\mu) &=& 
\frac{M C_0^2}{8\pi} 
\Biggl[ r_0  - \frac{g_A^2 M}{24 \pi
f_\pi^2} \Biggl( 3 - 8\frac{\mu}{m_\pi} + 6\frac{\mu^2}{m_\pi^2} 
     \nonumber \\[2pt]
 & &  \qquad\qquad\qquad\qquad\quad \null +
\frac{16\pi}{M m_\pi C_0} \left( 3\frac{\mu}{m_\pi} - 2\right) +
\frac{96\pi^2}{\left(Mm_\pi C_0 \right)^2} \Biggr) \Biggr]
      + \order(\mu)\, ,  \label{C20}
\ee
where
\beq
\mu_\pi \equiv \mu + \alpha_\pi M \ln \frac{\mu}{m_\pi} +
\frac{(\mu-m_\pi)^2}{\LambdaNN} \, ,
\eeq
and $\LambdaNN=16\pi f_\pi^2/g_A^2 M\simeq 300\,$MeV.
At $\mu=m_\pi$, these expressions reproduce
the matching condition quoted in Ref.~\cite{PDS}.
The expression for $C_0$ also coincides with the leading order result 
Eq.~(\ref{C0m1}) for this choice of $\mu$.

Solving the renormalization group equation for $C_0$
is another way to generate its $\mu$ dependence.
A calculation order-by-order
in the number of loops is not equivalent to solving it in powers of
momentum due to the nonperturbative nature of the power counting.
Expanding the amplitude, which is an observable,
in powers of momentum requires that the sum of all
graphs to a given order is $\mu$-independent. 
At subleading order, the $\mu$-dependent graphs are shown in
Fig.~\ref{rgeqn}, and applying $\mu \frac{\partial}{\partial\mu}$ to 
their explicit expressions from Ref.~\cite{PDS} 
results in the renormalization group equation
\beq
\mu \frac{\partial C_0}{\partial \mu} = \frac{M\mu}{4\pi} \left[ C_0^2
+ \left( \frac{\alpha_\pi M}{\mu} C_0^2 + \frac{g_A^2}{2f_\pi^2} C_0
\right) \frac1{1+2(\mu-m_\pi)/\LambdaNN} \right] 
      + \order(\mu) \, ,
\eeq
whose solution is the expression Eq.~(\ref{C00}), already obtained  from
matching.
The solution is consistent with the power counting for all $\mu$,
behaving like $C_0 \sim 1/ \mu$. 
This is in contrast to a renormalization group analysis based
on a loop expansion, whose solution is inconsistent with the power counting 
for $\mu\simgt\LambdaNN$ \cite{PDS}.
Although the above discussion was just for the \sing channel, 
the results also hold for the \trip channel 
since the pion contribution is identical in the two cases to this order.

\begin{figure}
\vspace{-.2in}
\begin{center}
\leavevmode
\vspace{-.5cm}
\hbox{
\hspace{-.5cm}
\epsfxsize=3.4in
\epsffile{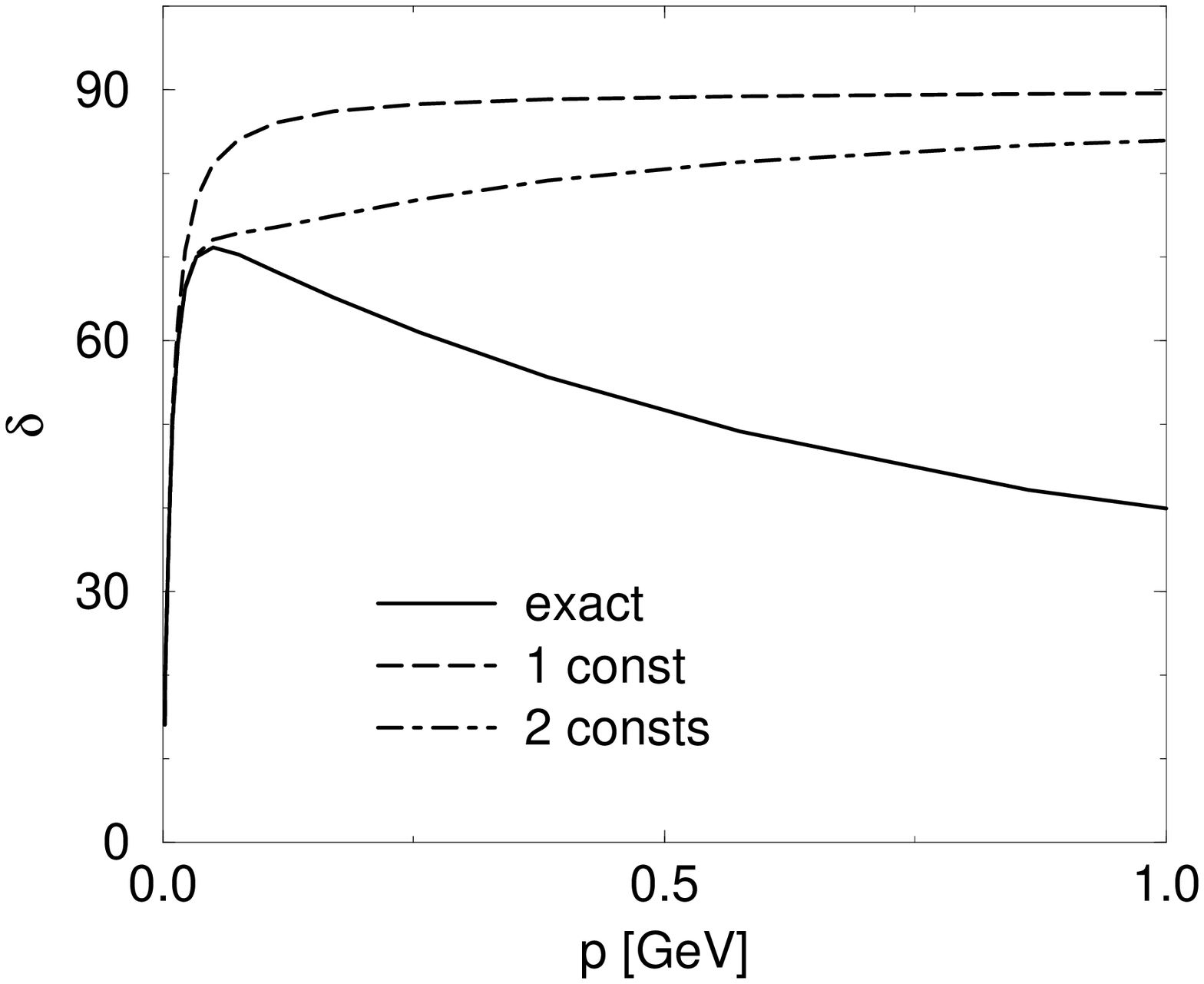}
\hspace{-.4cm}
\epsfxsize=3.4in
\epsffile{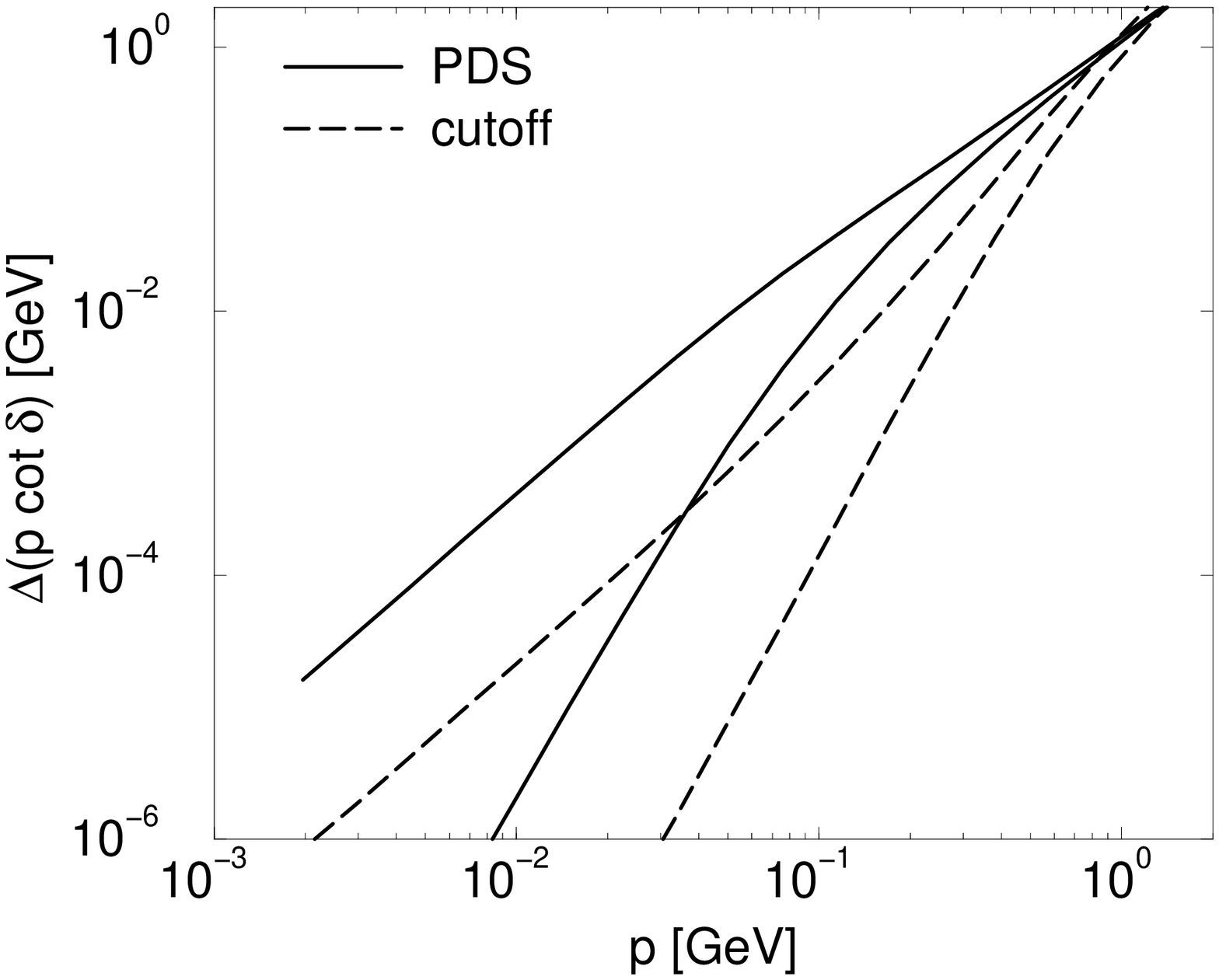}
}
\end{center}
\caption{\label{pds} The left plot shows the exact phase shift in
degrees for the
two-Yukawa model and the results from one (dashed) and two
(dot-dashed) constants in PDS with $\mu = m_\pi$.
The right side shows the error in $p\cot\delta$ for these cases
(solid) as well as for the cutoff regularization (dashed) of
Fig.~\protect\ref{yukawa}.}
\end{figure}

\subsection{PDS and the Modified Effective Range Expansion}

We can now apply the relations Eq.~(\ref{C0m1})--(\ref{C20}) to
generate the leading and subleading behavior of PDS.
The error plots for the two-Yukawa model from Sec.~II.C using 
PDS are compared in Fig.~\ref{pds}
with the cutoff results.
Matching the constants to data  ensures $\mu$-independent
observables only to the order they were fixed; the 
remaining error still depends on the
choice of $\mu$.
This is different from the case with a short-range potential only
\cite{OurPaper}, since  
pions contain $\mu$-dependent terms to all orders in the momentum
expansion. 
However, the results are relatively stable for $\mu\le450\,$MeV.
The radius of convergence (from extrapolating the slopes at small
$p$) is about $p\simeq m_\pi$, which is much smaller
than from cutoff regularization.%
\footnote{
There is a narrow
window  around $\mu=400\,$MeV in which the radius of convergence
increases to about 300~MeV.
Fine-tuning $\mu$ gives results qualitatively similar to
fine-tuning $D_2$ as a third parameter as done in Ref.~\cite{PDS}.
In that fit, however, the accuracy in the determination of the ER parameters
$a_s$ and $r_e$ was sacrificed in order to minimize  global errors
for $p \le 200\,$MeV.
In contrast,
Eqs.~(\ref{C00})--(\ref{C20}) can be used to fine-tune  $\mu$ while still
observing the matching requirements, leading to proper error plots and
systematic results. 
We will not exploit this effect but instead take $\mu = m_\pi$ as
in Ref.~\cite{PDS}.%
}
In fact, using a purely short-range effective potential with cutoff
regularization gives almost the same results as PDS with pions.

This implies the pion is not being properly taken into account,
as can be seen in the power counting of $C_2$.
Without pions, the scaling is $C_2\sim 1/\mu^2$, 
but Equation~(\ref{C20}) shows the pion contributions begin to
dominate for $\mu\sim m_\pi$, leading to 
\beq
C_2(\mu > m_\pi) \sim -\frac{8\pi\alpha_\pi}{m_\pi^2}\, .
\label{C2scale}
\eeq
The power counting for $C_2p^2$ therefore breaks down
for momenta above $m_\pi$. 
A nonperturbative treatment of pions might remedy this
situation. 
It is instructive, therefore, to investigate how the PDS power counting
differs from the modified ER expansion, which was so successful in the
previous section.

\begin{figure}[t]
\vspace{-.3in}
\begin{center}
\leavevmode
\epsfxsize=5in
\epsffile{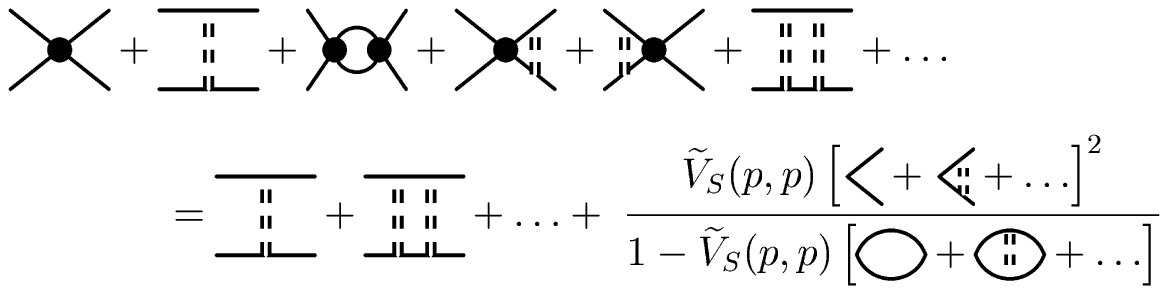}
\end{center}
\caption{\label{identity} A Feynman-diagram identity for the
Lippmann-Schwinger equation in dimensional regularization.}
\end{figure}

First, we show that, for dimensional regularization,
solving the Schr\"odinger equation with the
potential Eq.~(\ref{Veff1}) is the same as using the modified ER
expansion.
This can be shown diagrammatically for the truncated potential 
$V_S(p,q)=C_0+C_2(p^2+q^2)/2$.
The  amplitude ${\cal A}$ 
is found by solving the Lippmann-Schwinger equation, which is
equivalent to  solving the Schr\"odinger equation.
In dimensional regularization, factoring the potential 
out of the loop integrals produces an additional
contribution to the potential $\delta V_S(p,p)=-\alpha_\pi m_\pi M C_2$,
leading to $\wt{V}_S = V_S + \delta V_S$.
Denoting the long-range pion Yukawa%
\footnote{Note from
Eq.~(\ref{Veff1}) that part of the
pion contribution is proportional to a delta-function and is therefore
absorbed in $V_S$.} ($V_L$) with a double dashed line
and the short-range delta-function terms 
($V_S$) by a dot,  
the amplitude ${\cal A}$ can then be shown to
satisfy the Feynman-diagram identity \cite{KSW1}
in Fig.~\ref{identity}. 

The first group of terms on the right-hand side 
of Fig.~\ref{identity} is just the one-pion
exchange amplitude ${\cal A}_L$, which in turn is
related to the long-range phase shift $\delta_L$ by an expression
analogous to Eq.~(\ref{ampl}). 
The coefficient of $\wt{V}_S$ in the denominator of the last term is the
Green's function (see Appendix B) 
\beq
G_E(r=0,r'=0)=\langle  0| \frac1{E-H_L+i\epsilon}|0\rangle 
      = - \frac{M}{4\pi r}
\frac{f_L(p,r)}{{\cal F}_L(p)} \Bigg|_{r=0}\, ,
\eeq
and the coefficient of $\wt{V}_S$ in the numerator is given by the square
of \cite{KSW1}
\beq
\int\! \frac{d^3q}{(2\pi)^3} \langle q | (1 + G_E V_L) |
p \rangle = \frac1{{\cal F}_L(p)} \, .
\eeq
Therefore the diagrammatic identity can be rewritten as
\beq
\frac{4\pi}{M} \frac{|{\cal F}_L(p)|^2}
{p\cot(\delta-\delta_L)-ip} 
= - \frac{\wt{V}_S(p,p)}{1-\wt{V}_S(p,p) G_E(0,0)} \, .
\eeq
Calculating $G_E(0,0)$ in the PDS scheme with nonperturbative pions
(see Appendix B) leads to the modified ER function Eq.~(\ref{MERF}),
\be
K^{\rm PDS}(p) &=& - \frac{4\pi}{M\, \wt{V}_S(p,p)} - \mu
- \alpha_\pi M \ln \frac{\mu}{m_\pi}  - c_\pi
\label{MERFPDS}
\\
&=& - \frac1{\wt{a}_s} + \frac12 \wt{r}_0 p^2 + \ldots
\, ,
\nonumber
\ee
where $c_\pi$ and therefore $\wt{a}_s$ are both regularization scheme
dependent.
Matching the first two terms in the $p^2$ expansion gives
\be
\wt{C}_0(\mu) &=& -\frac{4\pi}{M{\cal D}} + \alpha_\pi m_\pi
\frac{2\pi\wt{r}_0}{{\cal D}^2}
\\
\wt{C}_2(\mu) &=& \frac{2\pi \wt{r}_0}{M {\cal D}^2}
\\
{\cal D} &=& \mu + \alpha_\pi M \ln \frac{\mu}{m_\pi} -
\left(\frac1{\wt{a}_s}-c_\pi\right)
\nonumber
\ee
where $\wt{C}_0$ includes the short-distance pion contribution
[see Eq.~(\ref{Veff1})].
These expressions are very different from the perturbative pion results
Eq.~(\ref{C00}-\ref{C20}) and lead to 
\beq
\wt{C}_2(\mu)  \sim \frac{2\pi\wt{r}_0}{M}  \frac{1}{\mu^2}\, ,
\eeq
which does not exhibit a breakdown in the scaling.

\begin{figure}
\begin{center}
\leavevmode
\hbox{
\hspace{-.5cm}
\epsfxsize=3.4in
\epsffile{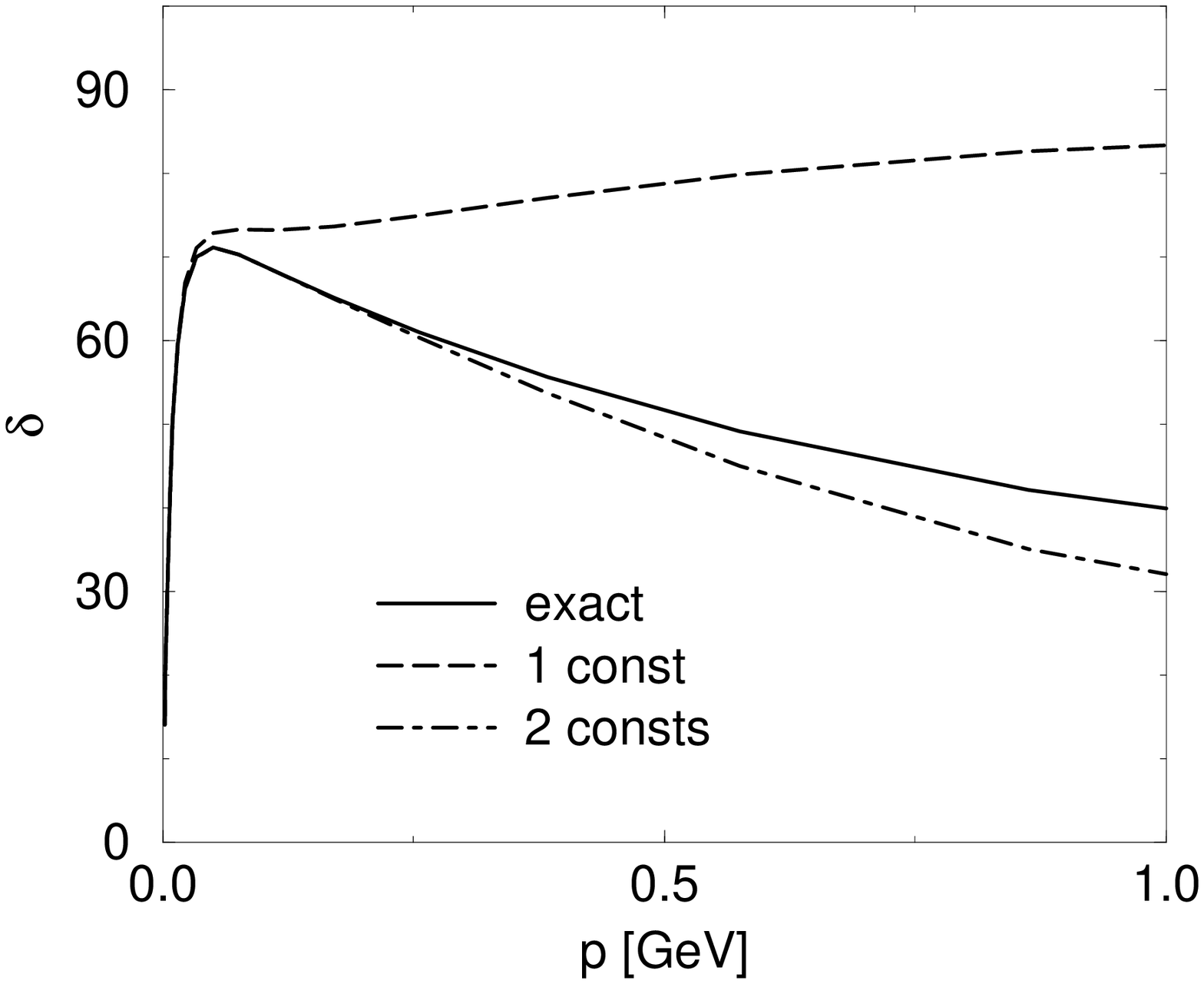}
\hspace{-.4cm}
\epsfxsize=3.4in
\epsffile{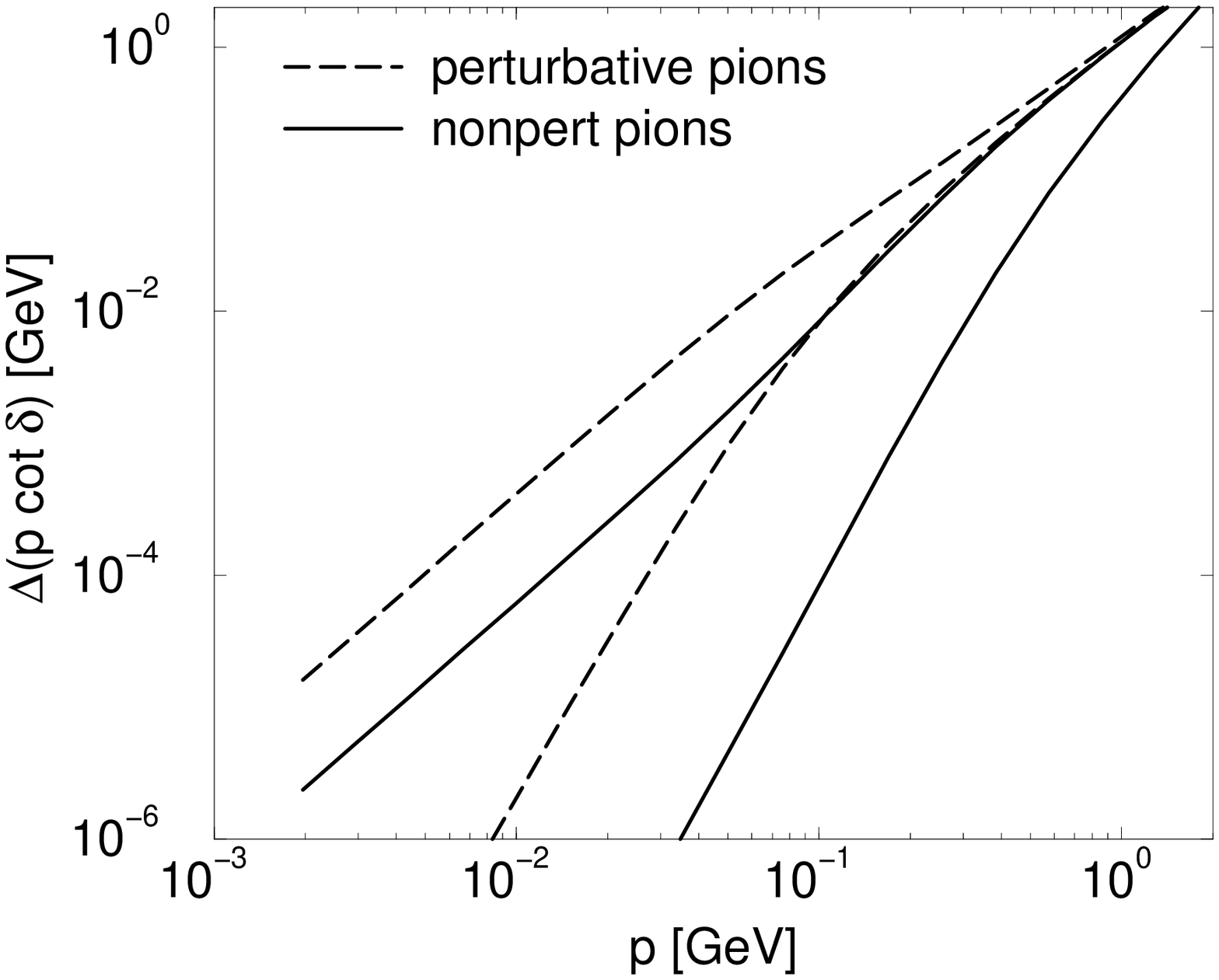}
}
\end{center}
\caption{\label{pds2} The left plot shows the exact phase shift in
degrees for the
two-Yukawa model and the results from one (dashed) and two
(dot-dashed) constants in PDS with nonperturbative pions.
The right side shows the error in $p\cot\delta$ for these cases
(solid) as well as for the perturbative pion case (dashed) of
Fig.~\protect\ref{pds}.}
\end{figure}

Removing the effect of the pions nonperturbatively, as in the above
modified ER 
procedure, leaves only a residual short-range potential.
PDS reproduces the modified ER expansion by construction in this case,
with $\mu$-independent results just as for a purely short-range
potential \cite{OurPaper}. 
The error plots for the two-Yukawa model, shown in Fig.~\ref{pds2},
illustrate the improvement of the PDS radius of convergence
from about $150\,$MeV with perturbative pions to
about $800\,$MeV with nonperturbative pions.

Taking pions into account perturbatively, as in the original
PDS power counting scheme, leads to a breakdown in the scaling of
the EFT constants.
This is reflected by the radius of convergence still set by $m_\pi$ as
in Fig.~\ref{pds}.
Implementing nonperturbative pions, either through using the modified ER
expansion or solving the Schr\"odinger equation,
rectifies the situation, leading to an improved radius of convergence
set by the next scale in the underlying theory.
It is not clear if some modification of the perturbative pion counting
scheme can be devised to transcend this problem.
For example, a global fitting procedure that does not enforce an exact
match of the low energy data could lead to a better overall error
\cite{PDS} in Fig.~\ref{pds2}.  However, the ability to determine a
radius of convergence graphically is lost in this case, requiring
further analysis 
to determine the breakdown scale.

\subsection{Cutoff Regularization and the Modified Effective Range Expansion}

We return to cutoff regularization within the context of the above
arguments. 
In this case, the potential $V_S$
cannot be factored out of the loop integrals in Fig.~\ref{identity},
so the Schr\"odinger equation and modified ER expansion are not as
simply related.
In particular,
the finite cutoff generates additional terms on the right-hand-side
of Fig.~\ref{identity} 
that contribute to the contact interactions.
For example, 
\beq
\parbox{10mm}{
\epsffile{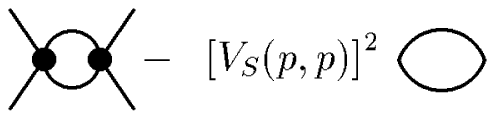}
}
\sim\;
-\frac{M\Lambda_c^3}{6\pi^2} 
     \Bigl({C}_0 + \frac{4\pi\alpha_\pi}{m_\pi^2}\Bigr) C_2
 + \ldots \, ,
  \label{lhsid}
\eeq
which can be interpreted as a shift of the short-range potential 
$V_S$ in the modified ER expansion.
The combination in parentheses originates from Eq.~(\ref{Veff1}).
It was shown in Ref.~\cite{OurPaper} that an unnatural cutoff leads
to an unnaturally large $C_0$, reducing
the radius of convergence of the EFT.
Thus an analogous effect should follow if
$\alpha_\pi$ is large.
This is why a strong long-range potential such as in the exponential and
delta-shell model of Sec.~II.A has better results in cutoff
regularization from using the modified ER expansion as opposed to the 
Schr\"odinger
equation.
However, a weak coupling, as occurs in NN scattering, 
leads to almost no difference in the radius of convergence.

\section{Application to Data}

Now that we have a better understanding of what sets the breakdown
scale for the EFT, we can apply the techniques of the previous
sections to the problem of NN scattering.
We take our data from the Nijmegen partial wave analysis
\cite{NNdata}, 
since it gives the most stable results over a large momentum range.
As we saw in the two-Yukawa model of Sec.~II.C, the ER
and modified ER parameters reflect the corresponding scales of the
underlying physics quite nicely.
Table~\ref{tab1} makes the same comparison using the $np$
scattering $S$-waves.%
\footnote{The precise values in Table~\ref{tab1} are 
somewhat sensitive to our choice of data and fitting procedure.} 
Note that the mass scale needed to make the modified
ranges $\wt{r}_n$ of order unity is actually {\it smaller} than the ranges
$r_n$, in stark contrast to the clean systematics of the two-Yukawa
model.

\begin{table}[t]
  \caption{\label{tab1}Conventional and modified effective range parameters
    for $S$-wave $np$ scattering.  The pion Yukawa was cut off at
    $p=m_\rho/2$.  All numbers are in appropriate powers of fm. 
     In parentheses the ranges are expressed in terms of 
    possible underlying scales, with 
    $m_\sigma=500\,$MeV.}
\vspace{.5cm}
\begin{tabular}{cc|ll@{\ }ll@{\ }ll@{\ }l}
& & \multicolumn{1}{c}{$a_s$ (fm)} & \multicolumn{2}{c}{$r_0$ (fm)}
& \multicolumn{2}{c}{$r_1/\Lambda^2$ (fm$^3$)}
& \multicolumn{2}{c}{$r_2/\Lambda^4$ (fm$^5$)} \\ \hline
 & ER & $-23.4$ & $2.63$ & ($1.9/m_\pi$) & $-0.25$ & ($-0.091/m_\pi^3$) &
$1.7$ & ($0.31/m_\pi^5$) \\
\raisebox{1.5ex}[0pt]{\sing}
 & Mod. ER & $-1.70$ & $3.10$ & ($7.9/m_\sigma$)
& $1.5$ & ($24./m_\sigma^3$)
& $-0.53$ & ($-55./m_\sigma^5$) \\ \hline
& ER & $5.39$ & $1.75$ & ($1.2/m_\pi$) & $0.21$ &
($0.074/m_\pi^3$) & $0.16$ & ($0.030/m_\pi^5$) \\
\raisebox{1.5ex}[0pt]{\trip}
&  Mod. ER & $-7.46$ & $2.12$ & ($5.4/m_\sigma$)
& $0.32$ & ($5.2/m_\sigma^3$)
& $0.28$ & ($29./m_\sigma^5$) \\
\end{tabular}
\end{table}

\begin{figure}
\begin{center}
\leavevmode
\hbox{
\hspace{-.5cm}
\epsfxsize=3.4in
\epsffile{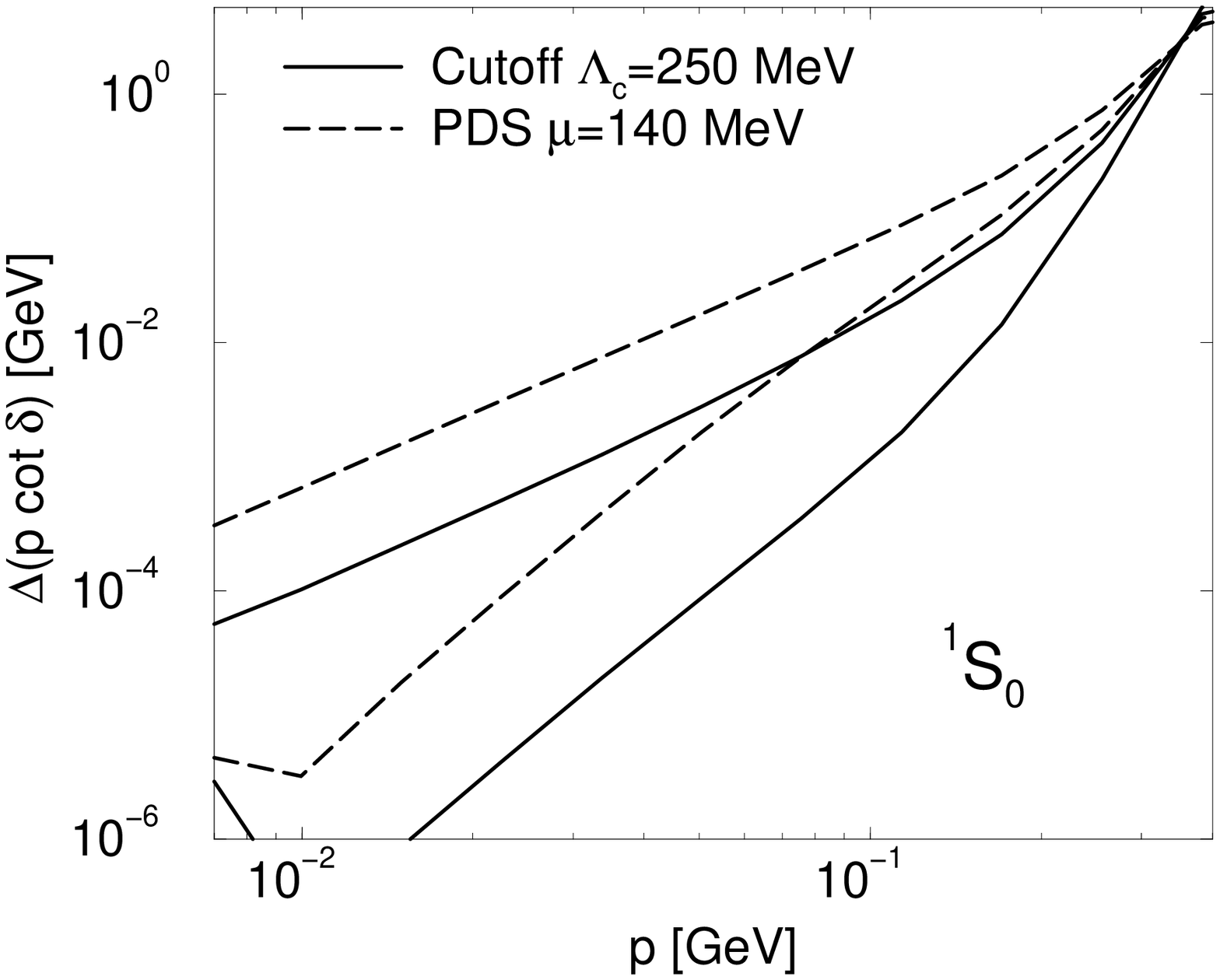}
\hspace{-.4cm}
\epsfxsize=3.4in
\epsffile{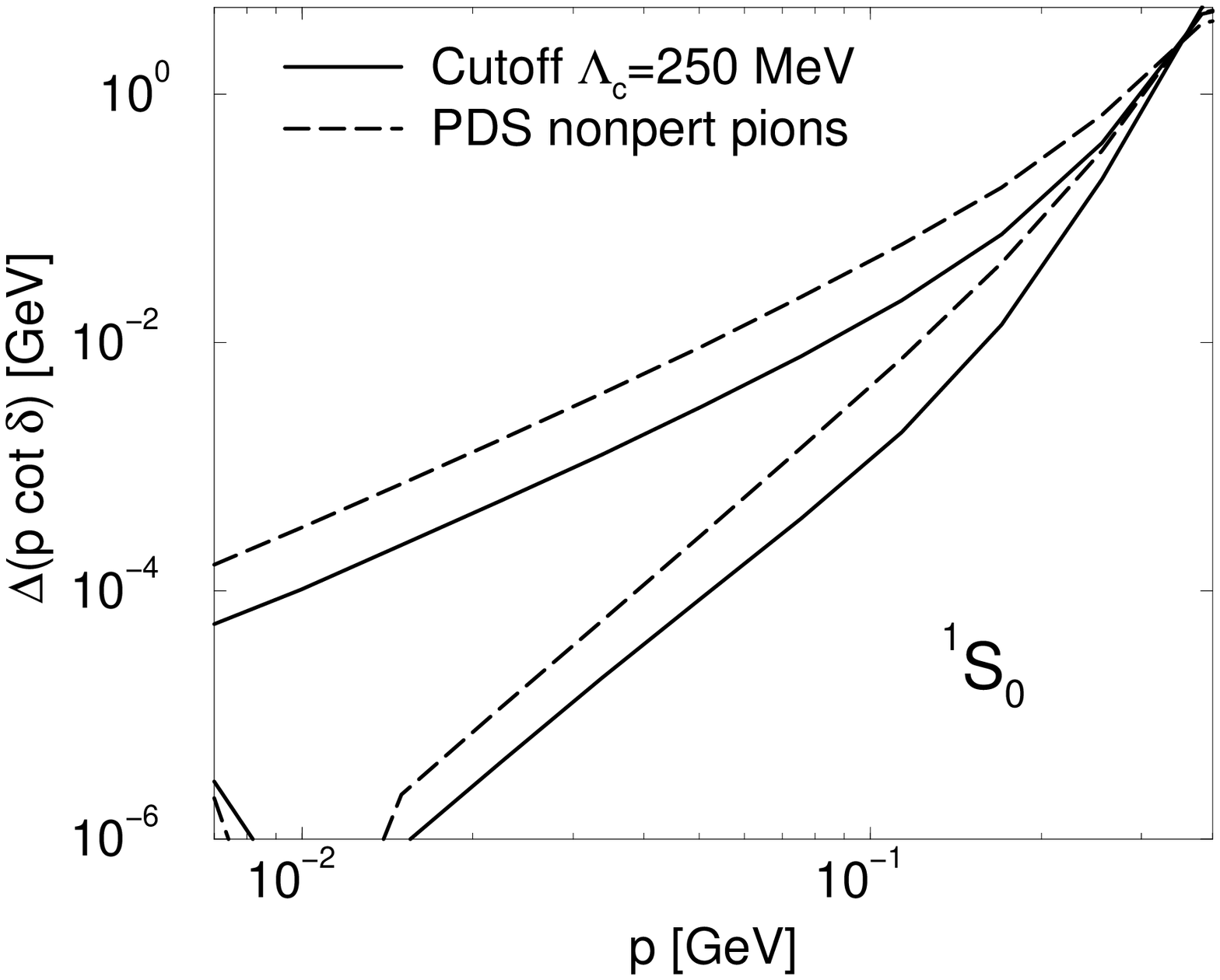}
}
\end{center}
\caption{\label{1S0} The \sing $np$ error in $p\cot\delta$ for
cutoff regularization with $\Lambda_c=250\,$MeV (solid) and for PDS
(dashed), both shown for one and two constants.
The PDS results are for $\mu = m_\pi$ and
perturbative pions in the left plot 
and for nonperturbative pions in the right plot.}
\end{figure}

This difference can also be seen in the error plots for the \sing
partial wave in Fig.~\ref{1S0}.
As anticipated by the two-Yukawa model, the modified ER and conventional ER
results for cutoff regularization give 
identical results, due to the weak coupling of the pion.
However, the optimal cutoff is
$\Lambda_c=250\,$MeV with a resulting radius of convergence
of about $300\,$MeV,
consistent with the analysis of Ref.~\cite{Lepage}.
This is much lower than the $500$-$600\,$MeV radius of convergence of
the two-Yukawa model. 
The PDS results for $\mu=m_\pi$ (left plot of Fig.~\ref{1S0})
break down earlier than cutoff
regularization with a worse absolute error,
although $\mu$ can again be tuned to improve the PDS error plots 
(as in Sec.~II.C).
Using PDS with nonperturbative pions improves the radius of
convergence (right plot of Fig.~\ref{1S0}), leading to results
comparable with the cutoff, as also seen in the last section.

The fact that the modified effective ranges and the radius of
convergence reflect a smaller scale than $500\,$MeV
must be due to a feature of the data that
our two-Yukawa model did not simulate. 
Retardation and relativistic
effects are not included in the model, but at these energies
they are small and should be negligible.
In the $np$ system, electromagnetic effects
are very small
in phenomenological models that parameterize the data \cite{NNdata,argonne}.
Details of the hadronic structure are only expected to appear for momenta
above $500\,$MeV \cite{Lepage}.
Errors in the value of $\alpha_\pi$ or in the data
have some effect on the values of the higher-order range parameters, 
but do not appreciably change the ranges quoted in
Table~\ref{tab1} or the error plot of Fig.~\ref{1S0}.  

There are two features which are not as easy to rule out:
the pion-production inelastic threshold,
at about $p\simeq360\,$MeV, and correlated two-pion exchange, which
gains strength above $p\simeq400\,$MeV,
both of which are close to the breakdown scale in Fig.~\ref{1S0}.
Conventionally, the latter effect has been modeled using a $\sigma$
meson with a mass of about $500\,$MeV \cite{machleidt}.  
Naive dimensional analysis shows that two-pion effects
are the same order as the $C_2$ contact term, and could prove to be
important \cite{Lepage}.
This would imply a breakdown at about $m_\sigma/2\sim 250\,$MeV 
and modified effective ranges with behavior $\wt{r}_n\sim1/m_\sigma$.
The error plots of the data (Fig.~\ref{1S0}) are consistent with this
effect, 
but the ranges, as shown in Table~\ref{tab1}, are not.

We can test the dependence of our analysis on two-pion
contributions by replacing the single short-distance Yukawa in the
two-Yukawa model by two short-distance Yukawas representing
 an attractive
$\sigma$ meson of mass $500\,$MeV and a repulsive vector meson 
(normally associated with both the $\omega$ and $\rho$ mesons, but we
will generically label it with a $\rho$):
\beq
V(r)\stackrel{r\to\infty}{\longrightarrow} \alpha_\rho
\frac{e^{-m_\rho r}}{r} - \alpha_\sigma \frac{e^{-m_\sigma r}}{r} -
\alpha_\pi \frac{e^{-m_\pi r}}{r} \, .
\label{threeyuk}
\eeq
Our two-Yukawa model Eq.~(\ref{twoyuk}) averaged the two short-distance 
effects into a single (attractive) short-ranged Yukawa for simplicity.
Choosing an arbitrary value for $\alpha_\sigma$, we tune $\alpha_\rho$
to obtain the large scattering length of NN scattering $S$-waves.
We consider two representative solutions: one with weak couplings
$(\alpha_\sigma,\alpha_\rho)=(1.5,1.5)$ and one with strong couplings
$(\alpha_\sigma,\alpha_\rho)=(7,14.65)$.
The strong couplings are similar to couplings used in the
Bonn potential to describe NN scattering \cite{machleidt} and are the
size expected from naive dimensional analysis \cite{fst}.

When the weak couplings are used, 
the modified ER approach provides a clean removal of the short-distance
physics in both the error plots and the $\wt{r}_n$'s, similar to
what was seen for the two-Yukawa model in Sec.~II.C. 
Applying the same procedure to the strong coupling
solution, however, gives remarkably similar results to those found with
the data.
In particular, the effective ranges behave similarly to those in
Table~\ref{tab1}, with the underlying scale for the modified effective
ranges seemingly worse than the conventional effective ranges, and the radius
of convergence drops from 
$600\,$MeV to $300\,$MeV in the error plots. 
In our model, we can explicitly prove that the breakdown is associated with
$m_\sigma/2$ (and not $2m_\pi$ for example) by increasing the scalar mass.
As $m_\sigma$ increases to $m_\rho$, the results smoothly shift to those of
the two-Yukawa model.

Therefore, the pion is successfully removed by the modified ER
expansion, leaving the next scale $m_\sigma/2$ to set the radius of
convergence.  
Just as a weak pion did not require the modified ER expansion
for an improved radius of convergence, a weak 
scalar coupling $\alpha_\sigma$ does not affect the breakdown.
This explains why only the strong coupling solution fails at
$m_\sigma/2$. 
A strong coupling also means the range of the interaction is not
$\order(1)\times 1/m_\sigma$ \cite{machleidt}.
The dimensionless strength of the potential is
$\eta_\sigma\equiv\alpha_\sigma
M/m_\sigma\sim13$.
Investigation of the three-Yukawa model Eq.~(\ref{threeyuk})
indicates that the modified effective ranges 
should scale as $\wt{r}_n\sim\eta^n_\sigma/m_\sigma$. 
The results in Table~\ref{tab1}
are indeed consistent with this scaling.

Thus the large modified effective ranges in NN scattering 
$\wt{r}_n$ imply that two-pion contributions are strong.
If this is so, these
two-pion effects should be included in the modified ER
procedure as long-distance physics in order to go beyond $p\sim300\,$MeV.
It is straightforward to test this conclusion in the three-Yukawa
model by including the $\sigma$ in the long-distance potential.
As expected, this leads to an improved radius of convergence.
The next step should be to use the explicit form for the two-pion
physics in an EFT analysis of data
with the modified ER procedure.

\section{Summary}

An effective field theory (EFT) approach to low-energy QCD 
systematically treats the physics of different length
scales.
For applications to nonrelativistic NN scattering 
and nuclear structure,
distance scales corresponding to momenta up to several times
the pion mass may be relevant.
In this regime,
the long-distance physics is dictated by the pion through chiral
symmetry.
The short-distance physics is replaced by a generic
expansion in powers of the momentum over a scale determined by the
remaining underlying interactions.
This conceptually simple procedure promises a systematic
description of two-nucleon processes.

The nonperturbative nature of nuclear forces complicates 
this EFT approach,
leading to open questions about including pions within various
regularization and power counting schemes.
In this paper,
we have examined cutoff regularization
and 
dimensional regularization with the PDS scheme. 
Toy models were used to analyze both approaches in a controlled
fashion before confronting NN scattering data.
We find that the implementation of these schemes 
can be subtle.

Within cutoff regularization, simply adding a one-pion-exchange
potential to a short-range effective potential does not necessarily
improve the radius of convergence of the EFT.
This is because the cutoff mixes long-distance contributions 
with the short-distance constants, which 
can lead to unnaturally large corrections.
Only if the long-distance coupling is weak, as occurs 
for one-pion exchange in
NN scattering, are these extra contributions negligible.
By using the modified effective range (ER) procedure, however,
a consistent
removal of the long-distance physics can be implemented for all
strengths of the interaction.

In the PDS scheme, the long-distance pion physics is accounted for
perturbatively with a consistent power counting prescription.
A systematic error analysis of a two-scale model showed, however,
that an EFT with the PDS power counting
breaks down earlier than cutoff regularization.
This breakdown is manifested by the renormalization-group
scaling of the short-distance constants.
The modified ER procedure was used to 
include one-pion exchange nonperturbatively in dimensional regularization
with PDS.
This removes
the long-distance contamination from the short-distance physics and
produces error plots
comparable to cutoff regularization.
It is not clear whether a consistent power counting with perturbative
pions can be 
developed to also include the benefits of this nonperturbative approach.

With the understanding of 
what causes the EFT breakdown in model problems, we 
applied the modified ER techniques to the analysis of NN scattering data.
Despite the clean removal of one-pion exchange effects, we found
 a radius of convergence of only $p\sim300\,$MeV and 
 unexpectedly large modified effective ranges.
We were only able to reproduce this characteristic behavior 
using a more realistic model of nuclear forces from one-boson-exchange
phenomenology.
This model includes strong 
attractive and repulsive Yukawa potentials 
resulting from the exchange of
scalar and vector mesons.
The mass of the scalar meson,
which is associated with correlated two-pion exchange, sets the
breakdown scale of the EFT, and the strong coupling is the source of
the large modified effective ranges.
This implies that to extend the radius of convergence beyond
$300\,$MeV, we must include correlated two-pion effects
explicitly in the EFT.
Since these effects are strong, a modified ER analysis could be essential.

\acknowledgments
We acknowledge useful discussions with J.~Gegelia, D.~B.~Kaplan,
M.~J.~Savage, B.~Serot, and N.~Tirfessa.
This work was supported in part by the National Science Foundation
under Grants No.\ PHY--9511923 and PHY--9258270.

\appendix

\section{Numerical Techniques}

The two quantities needed for numerical computation of the modified ER
are the Jost function ${\cal F}(p)$ and the derivative with respect to $r$ of
the Jost solution at the origin $f'(p,r)$.  
Both of these can be found by solving for the logarithmic derivative
of $f(p,r)$,
\beq
L_p(r)=\frac1{f(p,r)} \frac{\partial f(p,r)}{\partial r} \ .
\eeq
By differentiating $L_p(r)$ once and using the fact that $f(p,r)$ is a
solution to the Schr\"odinger equation, we arrive at the differential
equation
\beq
\frac{d L_p(r)}{dr}= 2m_{\rm red} V(r)-p^2-L_p^2(r),
\qquad\qquad
L_p(\infty)=ip,
\eeq
which can be solved as a coupled equation for both the real and
imaginary parts of $L_p(r)$ and for $L_p(0)\equiv L_p$ in particular.
The real part of $L_p$ is needed in the modified ER function,
Eq.~(\ref{MERF}), and the imaginary part can be shown to be
\beq
\Imag L_p= \frac{p}{|{\cal F}(p)|^2}\ ,
\eeq
which can be solved for the magnitude of the Jost function.

The phase of the Jost function is the negative of the phase shift for the
potential $V(r)$.
This can be found conveniently and accurately
using the variable phase method
\beq
\delta'(r)= - \frac{2m_{\rm red}}p V(r) \sin^2\left[ pr +\delta(r)
\right] \ ,
\qquad\qquad
\delta(0)=0 \ ,
\label{varph}
\eeq
with the full phase shift given by $\delta=\delta(\infty)$.
When numerically solving for a potential that includes a delta-shell,
\beq
V_\delta(r)= -\frac{\lambda}{2 m_{\rm red}} \delta(r-\rz) \ ,
\eeq
this method must be modified to take into account the discontinuity of
the wavefunction derivative at the radius $\rz$,
\beq
\phi'(\rz_+) - \phi'(\rz_-) = - \lambda \phi(\rz_-) \ .
\eeq
This can be accomplished by integrating Eq.~(\ref{varph}) out from the
origin to $r=\rz_-$ and then using the new value
\beq
\delta(\rz_+)= -p\rz + \arctan \left[
\frac{t(\rz_-)}{1-\frac{\lambda}p t(\rz_-)}\right] \ ,
\qquad
t(r)\equiv \tan\left(pr +\delta(r) \right)
\eeq
to continue the integration to infinity.

\section{Green's Function in the PDS Scheme}

We follow Ref.~\cite{KSW1} in solving the $S$-wave radial Schr\"odinger
equation.
Since the short-range potential Eq.~(\ref{Veff1}) is proportional to
a delta-function, we can separate it out to give
\beq
(H_L-E)\psi(r)=-V_S(r)\psi(r) =-V_S(r) \psi(0)\, .
\eeq
Away from the origin, the full wavefunction $\psi$ only
depends on  the regular and irregular solutions of the purely
long-range potential \cite{Newton}:
\be
\psi_R(r) &=& \frac{{\cal F}_L(-p)}{2ipr}  f_L(p,r)  + \mbox{c.c.}
\nonumber\\
\psi_I(r) &=& \left[\frac{1}{2{\cal F}_L(p)r} f_L(p,r) + \mbox{c.c.}
\right] - B\; \psi_R(r) \ ,
\label{solns}
\ee
with $B$ a momentum-dependent coefficient that is inconsequential to
our result.
These wavefunctions satisfy $(H_L-E)\psi_R=0$ and $(H_L-E)\psi_I=
(4\pi/M)\delta^3(r)$.
The Green's function 
$G_E(r,r'=0)=\langle 0|(E-H_L+i\epsilon)^{-1}|r\rangle$ 
can therefore be written as a
combination of the regular and irregular solution that cancels
the incoming spherical wave,
\beq
G_E(r,0)=-\frac{M}{4\pi} \left[ \psi_I(r)
 + \left( B+ \frac{ip}{|{\cal F}_L(p)|^2} \right) \psi_R(r) \right]
= -\frac{M}{4\pi r} \frac{f_L(p,r)}{{\cal F}_L(p)} \ ,
\label{green}
\eeq
with the last equality following from Eq.~(\ref{solns}).

The divergent quantity $G_E(0,0)$ can be separated into a momentum
dependent term needed for the modified ER function Eq.~(\ref{MERF}) and a
regularization scheme dependent constant
\be
- \frac{4\pi}{M} \Real G^{\rm reg}_E(0,0)
\equiv \Real \left[ \frac{f_L'(p,0)}{{\cal F}_L(p)}
\right]^{\rm reg}
- \frac{4\pi}{M} \Real G^{\rm reg}_0(0,0) \ .
\label{B4}
\ee
The first term is the logarithmic derivative of the Jost solution
$f_L(p,r)$ at the origin and, up to a constant, does not depend on the
regularization scheme.  Using the PDS scheme, 
the second term is \cite{KSW1,PDS}
\beq
-\frac{4\pi}{M}\Real G^{\rm PDS}_0(0,0) = \mu + \alpha_\pi M \ln
\frac{\mu}{m_\pi} + c_\pi \ .
\eeq
Since a momentum-independent term
can be shuffled between the two terms on the right-hand side of 
Eq.~(\ref{B4}), the constant $c_\pi$ is arbitrary. 
The imaginary part of $G_E(0,0)$ can be read directly from
Eq.~(\ref{green}), leading to the full Green's function
\beq
- \frac{4\pi}{M} G^{\rm PDS}_E(0,0) = 
\Real \left[ \frac{f_L'(p,0)}{{\cal F}_L(p)}
\right]^{\rm reg} + 
\mu + \alpha_\pi M \ln
\frac{\mu}{m_\pi} + c_\pi + \frac{ip}{|{\cal F}_L(p)|^2}\, ,
\eeq
as used in the text.

\end{document}